\documentclass[a4paper,11pt]{article}
\usepackage{jheppub} 

\usepackage{amsmath,amssymb,mathtools,slashed,xspace,graphicx,enumerate,xcolor,upgreek}
\RequirePackage[utf8]{inputenc}
\RequirePackage[british]{babel}

\usepackage{ifthen}
\usepackage{subcaption}
\usepackage{siunitx}
\usepackage{enumitem}
\usepackage{booktabs}
\usepackage{bookmark}
\usepackage{comment}

\usepackage{cleveref}
\newcommand{\mc}[1]{\mathcal{#1}}
\newcommand{\apollo}{{\sc Apollo}\xspace}
\newcommand{\alaric}{{\sc Alaric}\xspace}
\newcommand{\ariadne}{{\sc Ariadne}\xspace}
\newcommand{\dire}{{\sc Dire}\xspace}
\newcommand{\herwig}{{\sc Herwig}\xspace}
\newcommand{\sherpa}{{\sc Sherpa}\xspace}
\newcommand{\pythia}{{\sc Pythia}\xspace}
\newcommand{\vincia}{{\sc Vincia}\xspace}
\newcommand{\powheg}{{\sc Powheg}\xspace}
\newcommand{\mcatnlo}{{\sc MC@NLO}\xspace}

\newcommand{\alphas}{\ensuremath{\alpha_\mathrm{s}}}
\newcommand{\gs}{\ensuremath{g_\mathrm{s}}}
\newcommand{\NC}{\ensuremath{N_\mathrm{C}}}
\newcommand{\CA}{\ensuremath{C_\mathrm{A}}}
\newcommand{\CF}{\ensuremath{C_\mathrm{F}}}

\newcommand{\D}{\ensuremath{\mathrm{d}}}

\newcommand{\Wsct}{\ensuremath{W}^\mathrm{sct}}

\newcommand{\g}{\mathrm{g}}
\newcommand{\q}{\mathrm{q}}
\newcommand{\qbar}{\bar{\mathrm{q}}}
\renewcommand{\H}{\mathrm{H}}

\title{\boldmath A partitioned dipole-antenna shower with improved transverse recoil}

\author{Christian T Preuss}
\affiliation{Department of Physics, University of Wuppertal, 42119 Wuppertal, Germany}
\emailAdd{preuss@uni-wuppertal.de}

\abstract{
The implementation of a new final-state parton-shower algorithm in the \textsc{Pythia} event generator is described.
The shower algorithm, dubbed \textsc{Apollo}, combines central aspects of the \textsc{Vincia} antenna shower with the global transverse-recoil scheme of the \textsc{Alaric} framework in order to achieve formal consistency with next-to-leading logarithmic (NLL) resummation.
The shower algorithm is constructed in such a way that it facilitates a straightforward combination with fixed-order calculations.
As an explicit proof of concept, a general scheme for matrix-element corrections (MECs) and two separate multiplicative next-to-leading order (NLO) matching schemes are outlined. It is argued that both matching schemes retain the logarithmic accuracy of the shower.
The improved modelling of radiation is examined by contrasting the new algorithm with existing leading-logarithmic parton showers in \textsc{Pythia}.
}

\begin{document}
\maketitle

\section{Introduction}
Parton showers dress hard scattering events with soft and collinear radiation, describing the evolution from the high-energy parton level to the low-energy particle level, at which partons are confined into observable hadrons. Historically, parton-shower algorithms were formulated in terms of the Dokshitzer-Gribov-Lipatov-Altarelli-Parisi (DGLAP) splitting kernels
\cite{Gribov:1972ri,Altarelli:1977zs,Dokshitzer:1977sg}, which describe the factorisation of matrix elements in phase-space regions in which partons are emitted at small angle (collinearly) to their radiator.
Wide-angle soft gluon radiation can be incorporated into the DGLAP formalism by ordering emissions in decreasing angles \cite{Marchesini:1983bm}, leading to the notion of angular-ordered parton showers \cite{Platzer:2009jq}. 
When different evolution variables are chosen, soft coherence effects can similarly be restored through an explicit veto of emissions not ordered in angle  \cite{Bengtsson:1986et}.

Alternatively, shower branchings can be described as emissions from colour dipoles. This was first described in terms of ``Lund'' dipoles and implemented in the \ariadne shower \cite{Gustafson:1987rq,Lonnblad:1992tz}. In dipole-like shower algorithms, emissions are typically ordered in a notion of transverse momentum (often taken with respect to the emitting colour dipole) and shower branchings recoil locally against the parent dipole ends. 
The notion of a dipole shower can nowadays broadly be refined into so-called antenna showers \cite{Winter:2007ye,Giele:2007di,Giele:2011cb,Fischer:2016Vincia}, as implemented for instance in \vincia, and Catani-Seymour (CS) dipole showers \cite{Nagy:2005aa,Schumann:2007mg,Platzer:2011bc,Hoche:2015sya}, often simply referred to as dipole showers.
This distinction closely follows their analogies with fixed-order subtraction schemes, i.e., the antenna-subtraction scheme \cite{Kosower:1997zr,Campbell:1998nn,Gehrmann-DeRidder:2005btv,Daleo:2006xa,Currie:2013vh,Gehrmann:2023dxm} in the case of the former and the Catani-Seymour dipole-subtraction scheme \cite{Catani:1996jh,Catani:1996vz,Catani:2002hc} in the case of the latter.
Algorithmically, the two approaches differ in the decomposition of collinear- and soft-enhanced components of branching kernels across ``neighbouring'' dipoles and the absorbtion of the transverse recoil that is generated by the branching among the two dipole ends. While the antenna framework aims at an agnostic treatment of the emitting dipole ends, the dipole framework distinguishes between an emitting and a recoiling dipole end. In either case, the full singularity structure of squared tree-level matrix elements is restored upon summing over all dipoles. Another dipole-antenna-like approach is given by so-called sector showers \cite{Lopez-Villarejo:2011pwr,Brooks:2020upa}, in which the radiation phase space is decomposed into non-overlapping sectors, each of which correspond to a well-defined singularity structure of the respective squared matrix element. In sector showers, each branching kernel contains the full (unpartitioned) singularity structure of the respective sector.

Historically, angular-ordered showers have been known to reproduce next-to-leading logarithmic (NLL) resummation for global observables, due to their intrinsic connection to soft-gluon resummation \cite{Catani:1992ua}, but do not faithfully account for NLL effects in non-global observables \cite{Banfi:2006gy}.
For transverse-momentum-ordered (dipole) showers, on the other hand, it has been established that their local assignment of transverse recoil is inconsistent with NLL resummation of global observables \cite{Dasgupta:2018nvj,Nagy:2020rmk}. 
In \cite{Dasgupta:2020fwr}, two solutions have been proposed to overcome these limitations in dipole showers. Specifically, either changing the evolution variable to account for angular ordering while keeping the local recoil intact, or employing a global transverse-recoil scheme while keeping the evolution variable unchanged is generically consistent with NLL resummation for both global and non-global observables \cite{Dasgupta:2020fwr}.
Over the past years, substantial progress on the assessment \cite{Hoche:2017kst,Dasgupta:2018nvj,Nagy:2020rmk} and development 
\cite{Bewick:2019rbu,Dasgupta:2020fwr,Forshaw:2020wrq,Herren:2022jej} of logarithmically accurate parton showers has been made, including in particular
initial-state radiation \cite{Bewick:2021nhc,vanBeekveld:2022ukn,vanBeekveld:2022zhl,vanBeekveld:2023chs,Hoche:2024dee},
mass efects \cite{Assi:2023rbu},
matching to NLO calculations \cite{Hamilton:2023dwb,Assi:2023rbu}, and first steps beyond NLL accuracy \cite{FerrarioRavasio:2023kyg}.
These improvements are important steps towards high-precision simulation tools, especially in light of increasingly precise experimental measurements \cite{Campbell:2022qmc}.

Parton showers are only reliable in phase-space regions in which branchings occur at small relative energies or small angles to the radiator. 
In regions of phase space where hard, well-separated partons dominate, fixed-order calculations provide accurate predictions. 
An accurate modelling over the full phase space can be obtained through matching and merging schemes.
In (NLO) matching schemes, such as \mcatnlo \cite{Frixione:2002ik} and \powheg \cite{Norrbin:2000uu,Nason:2004rx,Frixione:2007vw}, a next-to-leading order (NLO) calculation is dressed with a parton-shower algorithm in such a way that mutual overlap in the real-radiation phase space is avoided. 
In (multi-jet) merging schemes, like matrix-element corrections (MECs) \cite{Bengtsson:1986et,Bengtsson:1986hr,Giele:2007di,Giele:2011cb,Lopez-Villarejo:2011pwr,Fischer:2016Vincia,Fischer:2017htu} or CKKW(-L) \cite{Catani:2001cc,Lonnblad:2001iq,Lonnblad:2011xx}, branching kernels are replaced by full (tree-level) matrix elements, either everywhere in phase space (MECs) or above a certain hardness scale (CKKW-L).
Matching and merging schemes have become essential parts of production-level simulations for modern collider experiments and often build on the exact factorisation of the branching phase space and the invertibility of the associated recoil scheme.

In the present work, a new parton-shower algorithm is constructed, borrowing key concepts from traditional antenna showers and combining them with the kinematics of the recently established \alaric shower \cite{Herren:2022jej,Assi:2023rbu}.
In particular, the shower evolution variable is defined as a notion of transverse momentum and branching probabilities are expressed in terms of antenna functions, suitably partitioned among the two antenna ends to facilitate the use of the global transverse-recoil scheme of \cite{Herren:2022jej,Assi:2023rbu}.
A guiding principle in the construction of the \apollo shower algorithm is to retain a close connection to fixed-order calculations, both in terms of matrix elements and in terms of the phase-space factorisation, so as to allow for the adaption of well-established matching and merging schemes.
While all concepts are in principal general, the discussion will be limited to the case of massless final-state partons, leaving an extension to initial-state radiation and massive partons to future work.
In reference to its implementation in the \pythia event generator \cite{Bierlich:2022pfr} and as an acronym of the fact that its \textbf{a}ntenna \textbf{p}artitioning \textbf{o}vercomes \textbf{l}ogarithmically \textbf{l}imiting \textbf{o}bstacles, the resulting shower algorithm is dubbed \apollo, the patron of the oracle of Delphi.

This manuscript is structured as follows. In \cref{sec:AntennaShowers}, the shortcomings of conventional antenna showers are briefly recapitulated before details of the \apollo shower algorithm are described for gluon emissions and gluon splittings separately. \Cref{sec:FixedOrderCorrections} describes the implementation of matrix-element corrections into the shower algorithm as well as two separate multiplicative NLO matching schemes. The consistency with NLL resummation is numerically tested for a wide range of event-shape observables in \cref{sec:ValidationResults}, which also contains a comparison to experimental data and existing parton showers in \pythia.
An outlook on future work is given in \cref{sec:Discussion}.

\section{Partitioned antenna showers}\label{sec:AntennaShowers}
In conventional antenna showers, the no-branching probability between two scales $t_n$ and $t'$ is constructed as
\begin{equation}
    \log\Delta_n(t_n,t') = -\sum\limits_{j}\int\limits_{t'}^{t_n} 8\uppi\alphas(t)\, \mathcal{C}_{ijk}\, A_{j/IK}(t,\zeta)\, \D t\, \D \zeta\, \frac{\D \phi}{2\uppi} \, ,
\label{eq:antennaSudakov}
\end{equation}
where the sum runs over all possible branchings $IK\mapsto ijk$ in (leading-colour) dipoles $IK$ and $\mc{C}_{ijk}$ is the associated (leading) colour factor.
The antenna functions $A_{j/IK}$ appearing in eq.~\eqref{eq:antennaSudakov} can be constructed directly from their single-unresolved limits as \cite{Braun-White:2023sgd}
\begin{align}
    A_{\g/\q\qbar}(p_i,p_j,p_k) &= \frac{2s_{ik}}{s_{ij}s_{jk}} + \frac{s_{jk}}{s_{ij}s_{ijk}} + \frac{s_{ij}}{s_{jk}s_{ijk}} \, , \label{eq:Aqgq}\\
    A_{\g/\q\g}(p_i,p_j,p_k) &= \frac{2s_{ik}}{s_{ij}s_{jk}} + \frac{s_{jk}}{s_{ij}s_{ijk}} + \frac{s_{ij}s_{ik}}{s_{jk}s_{ijk}^2} \, , \label{eq:Aqgg}\\
    A_{\g/\g\g}(p_i,p_j,p_k) &= \frac{2s_{ik}}{s_{ij}s_{jk}} + \frac{s_{jk}s_{ik}}{s_{ij}s_{ijk}^2} + \frac{s_{ij}s_{ik}}{s_{jk}s_{ijk}^2} \label{eq:Aggg}\, , \\
    A_{\q/\g X}(p_i,p_j,p_k) &= \frac{1}{s_{ij}} - \frac{2s_{jk}s_{ik}}{s_{ij}s_{ijk}^2} \label{eq:AqgX}\, .
\end{align}
In particular, these functions do not depend on the azimuthal angle $\phi$.
The evolution variable $t$ is typically chosen as a notion of transverse momentum known as \ariadne-$p_\mathrm{T}$, which in the case of massless particles is given by \cite{Gustafson:1987rq,Lonnblad:1992tz}
\begin{equation}
    t \equiv p_\mathrm{T}^2 = \frac{s_{ij}s_{jk}}{s_{ijk}} \, .
\label{eq:pTariadne}
\end{equation}
The auxiliary phase-space variable $\zeta$ can, for instance, be chosen as the rapidity with respect to the antenna parents,
\begin{equation}
    \zeta = \frac{1}{2}\log\left(\frac{s_{jk}}{s_{ij}}\right) \, .
\end{equation}
In the $j\parallel i$ limit, the \ariadne-$p_\mathrm{T}$ reduces to $p_\mathrm{T}^2 \to (1-z_i)s_{ij}$ with $z_i$ the collinear momentum fraction of particle $i$. Similarly, it reduces to $p_\mathrm{T}^2 \to (1-z_k)s_{jk}$ in the $j\parallel k$ limit with $z_k$ denoting the collinear momentum fraction of particle $k$.

For massless particles, the post-branching kinematics in antenna showers are constructed as
\begin{equation}
\begin{split}
    p_i^\mu &= a_i p_I^\mu + b_i p_K^\mu - c p_\perp^\mu \, ,\\
    p_j^\mu &= (1-a_i-a_k)p_I^\mu + (1-b_i-b_k)p_K^\mu + p_\perp^\mu \, , \\
    p_k^\mu &= a_k p_I^\mu + b_k p_K^\mu - (1-c) p_\perp^\mu \, ,
\end{split}
\label{eq:antennaKinematics}
\end{equation}
so that the transverse momentum 
\begin{equation}
    \vec{p}_\perp^2 = -p_\perp^2 = (1-a_i-a_k)(1-b_i-b_k)s_{ijk}
\end{equation}
generated by the emission of particle $j$ is absorbed locally in the dipole by the dipole ends $I$ and $K$. The parameters of the map are given by
\begin{equation}
\begin{split}
    a_i &= \frac{(1-y_{jk})(1-y_{jk}+\rho) - c(y_{ij}-y_{jk} + y_{ij}y_{jk} + y_{jk}^2)}{2\rho} \, ,\\        
    a_k &= \frac{-(1-y_{ij})(1-y_{jk}-\rho) - c(y_{jk}-y_{ij}+y_{ij}y_{jk}+y_{ij}^2) + 2y_{ij}y_{jk}}{2\rho} \, ,\\
    b_i &= \frac{-(1-y_{jk})(1-y_{jk}-\rho) + c(y_{ij}-y_{jk}+y_{ij}y_{jk}+y_{jk}^2)}{2\rho} \, ,\\
    b_k &= \frac{(1-y_{ij})(1-y_{jk}+\rho) + c(y_{jk}-y_{ij}+y_{ij}y_{jk}+y_{ij}^2) - 2y_{ij}y_{jk}}{2\rho} \, ,
\end{split}
\end{equation}
where $y_{ij} = s_{ij}/s_{ijk}$ and
\begin{equation}
    \rho = \sqrt{(1-y_{jk})^2 + c^2(y_{ij}+y_{jk})^2 - 2c(y_{ij}-y_{jk}+y_{ij}y_{jk}+y_{jk}^2)} \, .
\end{equation}
The parameter $c$ is in principle arbitrary as long as collinear safety in the $j\parallel i$ and $j\parallel k$ limits is guaranteed. This ensures that the transverse recoil is absorbed entirely by $p_I$ in the $j\parallel i$ limit ($c \to 1$) and by $p_K$ in the $j\parallel k$ limit ($c\to 0$). 
A practicable choice that can be constructed directly in the event frame is
\begin{equation}
    c = \frac{s_{jk}}{s_{ij}+s_{jk}} \, ,
\end{equation}
but other choices are possible, see e.g.~\cite{Gustafson:1987rq,Lonnblad:1992tz,Kosower:1997zr,Giele:2007di,Gehrmann-DeRidder:2011gkt}. Another special choice worth mentioning is $c=1$, corresponding to the kinematics in conventional dipole showers \cite{Sjostrand:2004ef,Nagy:2005aa,Dinsdale:2007mf,Schumann:2007mg,Platzer:2011bc,Hoche:2015sya}, in which $a_k=0$ and the transverse recoil is absorbed solely by $p_i$.

Ignoring for now issues in the assignment of subleading-colour factors in dipole showers, cf.~e.g.~\cite{Gustafson:1987rq,Giele:2007di,Giele:2011cb,Nagy:2012bt,Dasgupta:2018nvj,Hamilton:2020rcu}, it has been identified in \cite{Dasgupta:2018nvj,Dasgupta:2020fwr} that the assignment of the transverse recoil in dipole-like showers affects their ability to generically reach higher logarithmic accuracy than LL for global observables. 
Regardless of the specific choice of $c$, the form of the kinematics eq.~\eqref{eq:antennaKinematics} obstructs conventional antenna showers from reaching a formal accuracy beyond LL. This can be understood by following the arguments of \cite{Dasgupta:2018nvj} and considering a gluon emission from the $\g_j-\g_k$ antenna in a (leading-colour) configuration $\q_i-\g_j-\g_k-\qbar_k$. This corresponds to an emission on the ``back'' of the Lund plane. For any ordering of the emissions, the transverse recoil is always absorbed by $\g_j$ and $\g_k$, affecting their respective transverse momenta, even when the emission is far away from previous ones in one direction of the Lund plane. 

A straightforward attempt to overcome these limitations faced with conventional antenna kinematics is to let the antenna parents $I$ and $K$ recoil longitudinally
\begin{equation}
\begin{split}
    p_i^\mu &= ra_i p_I^\mu \, ,\\
    p_j^\mu &= r(1-a_i)p_i^\mu + r(1-b_k)p_k^\mu + rp_\perp^\mu \, ,\\
    p_k^\mu &= rb_k p_K^\mu \, ,
\end{split}
\label{eq:antennaKinematicsGlobal}
\end{equation}
while absorbing the transverse recoil generated by $p_j^\mu$ globally across the final-state particles,
\begin{equation}
    p_r^\mu = p_R^\mu - p_\perp^\mu = \sum\limits_{n\in \text{FS}} p_n^\mu - p_\perp^\mu \, .
\end{equation}
It is worth highlighting that the mapping \cref{eq:antennaKinematicsGlobal} changes the invariant mass of the antenna $s_{IK}$ to $s_{ijk} = s_{IK} - \vec{p}_\perp^2$ and of the entire event, $p_r^2 \neq p_R^2$. 
The latter is restored by rescaling all post-branching final-state momenta, including $p_i^\mu$, $p_j^\mu$, and $p_k^\mu$, by a factor
\begin{equation}
    r=\sqrt{\frac{p_R^2}{p_r^2}} \, .
\end{equation}
This approach is adopted in the PanGlobal antenna shower and has been demonstrated to satisfy NLL requirements \cite{Dasgupta:2020fwr}. 

In the following, an alternative approach will be described, resting on a judicious partitioning of antenna functions across collinear sectors \cite{Catani:1996vz}. The partitioned antenna functions allow gluon emissions to be generated with dipole-like branching kinematics in which the transverse recoil is absorbed globally \cite{Herren:2022jej}, therefore overcoming shortcomings of traditional antenna showers with respect to their ability to achieve consistency with NLL resummation.

This section is structured as follows. In \cref{sec:GluonRadiation} a framework for gluon emissions with formal NLL accuracy is discussed, combining central elements of traditional antenna showers \cite{Giele:2007di,Giele:2011cb} with the kinematics of the \alaric parton shower \cite{Herren:2022jej,Assi:2023rbu}, whereas \cref{sec:GluonSplittings} introduces gluon splittings into the \apollo framework. \Cref{sec:SectorShowers} discusses an extension of the new shower algorithm to so-called sector showers, cf.~\cite{Lopez-Villarejo:2011pwr,Brooks:2020upa,Brooks:2020mab}.

\subsection{Gluon radiation}\label{sec:GluonRadiation}
Introducing an auxilary vector $n_j^\mu$, with $p_in_j \neq 0$ and $p_kn_j \neq 0$, the soft eikonal can be partitioned as follows \cite{Catani:1996vz}
\begin{align}
    \frac{2s_{ik}}{s_{ij}s_{jk}} &= \frac{2s_{ik}}{s_{ij}s_{jk}}\frac{s_{jk}(p_in_j)+s_{ij}(p_kn_j)}{s_{jk}(p_in_j)+s_{ij}(p_kn_j)} \nonumber \\
    &= \frac{1}{s_{ij}}\frac{2s_{ik}(p_in_j)}{s_{jk}(p_in_j)+s_{ij}(p_kn_j)} + \frac{1}{s_{jk}}\frac{2s_{ik}(p_kn_j)}{s_{jk}(p_in_j)+s_{ij}(p_kn_j)} \, ,
\end{align}
leading to the following partitioned dipole-antenna functions
\begin{align}
    P_{\q\g}(p_i,p_j,p_k;n_j) &= \frac{1}{s_{ij}}\left[\frac{2s_{ik}(p_in_j)}{s_{jk}(p_in_j)+s_{ij}(p_kn_j)} + \frac{s_{jk}}{s_{ijk}} \right]\, , \label{eq:Pqg} \\
    P_{\g\g}(p_i,p_j,p_k;n_j) &= \frac{1}{s_{ij}}\left[\frac{2s_{ik}(p_in_j)}{s_{jk}(p_in_j)+s_{ij}(p_kn_j)} + \frac{s_{jk}s_{ik}}{s_{ijk}^2} \right]\, . \label{eq:Pgg}
\end{align}
The full antenna functions are recovered by symmetrising over the antenna parents $i$ and $k$,
\begin{align}
    A_{\g/\q\qbar}(p_i,p_j,p_k) &= P_{\q\g}(p_i,p_j,p_k;n_j) + P_{\q\g}(p_k,p_i,p_j;n_j) \, , \\
    A_{\g/\q\g}(p_i,p_j,p_k) &=  P_{\q\g}(p_i,p_j,p_k;n_j) + P_{\g\g}(p_k,p_i,p_j;n_j)\, , \\
    A_{\g/\g\g}(p_i,p_j,p_k) &= P_{\g\g}(p_i,p_j,p_k;n_j) + P_{\g\g}(p_k,p_i,p_j;n_j) \, .
\end{align}

A naive choice of $n_j^\mu$ would be $n^\mu = p_i^\mu + p_k^\mu$, which recovers the traditional Catani-Seymour partitioning of the eikonal \cite{Catani:1996vz},
\begin{equation}
    \frac{2s_{ik}}{s_{ij}s_{jk}} = \frac{1}{s_{ij}}\frac{2s_{ik}}{s_{ij}+s_{jk}} + \frac{1}{s_{jk}}\frac{2s_{ik}}{s_{ij}+s_{jk}} \,.
\end{equation}
In this case, the transverse recoil of the emission would typically be taken entirely by the emitter, corresponding to a choice of $c=1$ in \cref{eq:antennaKinematics}. This leads to correlations among multiple gluon emissions in the shower evolution, and therefore fails requirements on NLL-consistent parton showers.

Alternatively, one may construct the reference vector globally in the event and choose $n_j^\mu = K^\mu + p_j^\mu$, similar in spirit to the \alaric construction \cite{Herren:2022jej}, in turn based on the dipole-subtraction scheme with many identified hadrons \cite{Catani:1996vz},
\begin{equation}
    K^\mu = -\sum\limits_{k \in \text{FS}} p_k^\mu = \sum\limits_{k\in \text{IS}} p_k^\mu \, ,
\end{equation}
where the sum on the right-hand side of the second equality may either run over the two incoming partons in a scattering or the single momentum of a decayed resonance.
This allows to define a variable $z$ which in the $i\parallel j$ limit reduces to the collinear momentum fraction,
\begin{equation}
    z = \frac{p_in_j}{(p_i+p_j)n_j} \, .
\label{eq:zDefEmit}
\end{equation}
In terms of $z$, the branching kinematics $\tilde{p}_{ij}^\mu + \tilde{K}^\mu = p_i^\mu + p_j^\mu + K^\mu$ can be constructed in the event frame so that the emitter $\widetilde{ij}$ recoils purely longitudinally
\begin{equation}
\begin{split}
    p_i^\mu &= z \tilde{p}_{ij}^\mu \, , \\
    p_j^\mu &= a\tilde{p}_{ij}^\mu + b\tilde{K}^\mu + p_\perp^\mu\, , \\
    K^\mu &= (1-z-a)\tilde{p}_{ij}^\mu + (1-b)\tilde{K}^\mu - p_\perp^\mu \, ,
\end{split}
\label{eq:radiationKinematics}
\end{equation}
with a subsequent boost of all momenta in the event, including $p_i^\mu$ and $p_j^\mu$, into the rest frame of $K^\mu$ \cite{Herren:2022jej}.
The kinematical variables used in eq.~\eqref{eq:radiationKinematics} are uniquely determined by requiring an exact and collinearly safe phase-space factorisation with on-shell particles as\footnote{Note that the parameter $b$ is identified as the scaled virtuality $v$ in \alaric.}
\begin{equation}
    b = \frac{1}{z}\frac{s_{ij}}{2\tilde{p}_{ij}\tilde{K}} \, , \quad a = (1-b)(1-z)-2b\kappa \, , \quad \kappa = \frac{\tilde{K}^2}{2\tilde{p}_{ij}\tilde{K}} \, .
\end{equation}
The transverse recoil
\begin{equation}
    \vec p_\perp^2 = b((1-b)(1-z)-b\kappa)2\tilde{p}_{ij}\tilde{K}
\end{equation}
generated by the emission of particle $j$ is balanced by a cumulative recoil of the entire event $\tilde{K}$. As analysed in \cite{Herren:2022jej}, this is paramount to ensure NLL safety of the kinematics mapping.

The shower evolution variable is defined in analogy to the collinear limit of the \ariadne-$p_\mathrm{T}$ in eq.~\eqref{eq:pTariadne},
\begin{equation}
    t = (1-z)s_{ij} = z(1-z)2\tilde{p}_{ij}\tilde{K} \, , \label{eq:showerEvolEmit}
\end{equation}
and the auxiliary phase-space variable is conveniently chosen as $z$.
In terms of the shower variables $t$ and $z$, the one-particle phase space for a final-state gluon emission with final-state spectator\footnote{Note the difference between spectator momentum $\tilde{p}_k$ and recoiling momentum $\tilde{K}$, see also \cite{Herren:2022jej}.} is given by
\begin{equation}
    \D\Phi^\mathrm{FF}_\mathrm{branch}(t,z,\phi) = \frac{1}{16\uppi^2}\, \D t\, \frac{\D z}{1-z}\, \frac{\D \phi}{2\uppi}
\end{equation}

For the sake of implementing the branching kernels \cref{eq:Pqg,eq:Pgg} numerically in a veto algorithm, they can be overestimated by
\begin{equation}
    P^\mathrm{trial}_\mathrm{emit}(t,z) = \frac{5}{t} \, ,
\label{eq:PtrialEmit}
\end{equation}
so that the trial gluon-emission probability is given by
\begin{equation}
    \frac{\D \mc{P}^\mathrm{trial}_\mathrm{emit}(t)}{\D t} = \frac{\alphas(t)}{2\uppi}\,\frac{1}{t} \log\left(\frac{1-z_-}{1-z_+}\right)\, .
\end{equation}
An overestimate of the lower and upper bound of the $z$-integration can be found in terms of the lower bound on the evolution scale $t_0$ imposed by the shower cutoff and quark-mass thresholds as
\begin{equation}
     z_\mp = \frac{1}{2}\left(1\mp\sqrt{1-\frac{4t_0}{2\tilde{p}_{ij}\tilde{K}}}\right)
\end{equation}
Both $t$ and $z$ can then be generated using the veto algorithm.

\subsection{Gluon splittings}\label{sec:GluonSplittings}
A gluon-splitting antenna function can be constructed directly from its collinear behaviour as \cite{Braun-White:2023sgd}
\begin{equation}
    P_{\q\qbar}(p_i,p_j,p_k) = \frac{1}{s_{ij}}\left[1 - \frac{2s_{jk}s_{ik}}{s_{ijk}^2} \right]\label{eq:Pqq}\, .
\end{equation}
Typically, gluon splittings do not fit nicely into the antenna picture, due to the intrinsic asymmetry between the splitting and the recoiling antenna leg. Moreover, gluon splittings do not interfere with the logarithmic structure at NLL \cite{Herren:2022jej,Assi:2023rbu}, since they are driven by purely collinear splitting functions \cite{Banfi:2004yd}. It is therefore possible to exploit this freedom to define kinematics that exhibit a symmetry between the quark-antiquark pair in the splitting and without singling out another particle in the event as the sole recoiler.
Such a momentum map can be constructed by balancing the transverse recoil between the quark-antiquark pair and distributing the longitudinal recoil across the entire event. Specifically, the post-branching momenta in the splitting kinematics are given by \cite{Assi:2023rbu}
\begin{equation}
\begin{split}
    p_i^\mu &= a_i\tilde{p}_i^\mu + b_i\tilde{K}^\mu + p_\perp^\mu \, , \\
    p_j^\mu &= a_j\tilde{p}_i^\mu + b_j\tilde{K}^\mu - p_\perp^\mu \, ,\\
    K^\mu &= (1-a_i-a_j)\tilde{p}_{ij} + (1-b_i-b_j)\tilde{K}^\mu \, ,
\end{split}
\label{eq:splittingKinematics}
\end{equation}
with a subsequent boost into the rest frame of $K^\mu$. 
The parameters in \cref{eq:splittingKinematics} are uniquely defined upon requiring a collinearly safe map with on-shell post-branching momenta. Explicitly, they are given by
\begin{equation}
\begin{split}
    a_i &= \frac{1}{2(1+\kappa)}\left(\bar{z}+(1-\bar{z})y+(1+2\kappa)\frac{(1-y)(y-\bar{z}(1+y))+2y\kappa}{\sqrt{(1-y)^2-4y\kappa}} \right) \, ,\\
    a_j &= \frac{1}{2(1+\kappa)}\left((1-\bar{z})+\bar{z}y-(1+2\kappa)\frac{(1-y)(1-\bar{z}(1+y))-2y\kappa}{\sqrt{(1-y)^2-4y\kappa}}\right) \, ,\\
    b_i &= \frac{1}{2(1+\kappa)}\left(\bar{z}+(1-\bar{z})y-\frac{(1-y)(y-\bar{z}(1+y))+2y\kappa}{\sqrt{(1-y)^2-4y\kappa}} \right) \, ,\\
    b_j &= \frac{1}{2(1+\kappa)}\left((1-\bar{z})+\bar{z}y+\frac{(1-y)(1-\bar{z}(1+y))-2y\kappa}{\sqrt{(1-y)^2-4y\kappa}}\right) \, ,
\end{split}
\end{equation}
where the phase-space variables $y$ and $\bar{z}$ are defined as
\begin{equation}
    y = \frac{p_ip_j}{p_ip_j + (p_i+p_j)K} \, , \quad \bar{z} = \frac{p_iK}{(p_i+p_j)K} \, .
\label{eq:splittingKinVariables}
\end{equation}
Note that \cref{eq:splittingKinematics} implies $s_{ij} = y (2\tilde{p}_{ij}\tilde{K})$. By construction, the transverse momentum
\begin{equation}
    \vec{p}_\perp^2 = \frac{y(\bar{z}(1-\bar{z})(1-y)^2-y\kappa)}{(1-y)^2-4y\kappa}2\tilde{p}_{ij}\tilde{K} \, ,
\end{equation}
generated through the splitting of particle $\widetilde{ij}$ is balanced between the post-branching particles $i$ and $j$. In particular, this implies that this map does not satisfy NLL safety requirements when combined with a transverse-momentum ordered shower evolution. Since these kinematics are employed only for gluon splittings, which do not contain a single-soft enhancement, this does not interfere with the NLL accuracy of the shower algorithm.

The evolution variable for gluon splittings is defined identically to \cref{eq:showerEvolEmit} as
\begin{equation}
    t = (1-z)s_{ij} = y(1-y)(1-\bar{z}) 2\tilde{p}_{ij}\tilde{K} \, . \label{eq:showerEvolSplit}
\end{equation}
For consistency reasons and to simplify multiplicative matching algorithms, cf.~\cref{sec:NLOMatching}, the auxiliary phase-space variable is chosen as in the gluon-emission case,
\begin{equation}
    z = \frac{p_in_j}{(p_i+p_j)n_j} = \bar{z}(1-y)+y \, ,
\end{equation}
which approaches the collinear momentum fraction in the collinear limit.
The one-particle phase space for a final-state splitting can then be expressed in terms of $t$ and $z$ as
\begin{equation}
    \D\Phi^{\mathrm{FF}}_\mathrm{split}(t,z,\phi) = \frac{1}{16\uppi^2}\, \D t\, \frac{\D z}{1-z}\, \frac{\D \phi}{2\uppi} \, .
\end{equation}

For practical purposes, the branching kernel \cref{eq:Pqq} can be overestimated by
\begin{equation}
    P^\mathrm{trial}_\mathrm{split}(t,z) = \frac{2(1-z)}{t} \, ,
\label{eq:PtrialSplit}
\end{equation}
With an overestimated $z$ range of $z_- =0$ and $z_+=1$, the trial gluon-splitting probability is then given by
\begin{equation}
    \frac{\D \mc{P}^\mathrm{trial}_\mathrm{split}(t)}{\D t} = \frac{\alphas(t)}{2\uppi}\,\frac{2}{t} \, ,
\end{equation}
and both $t$ and $z$ can be generated using standard Monte-Carlo techniques.

\subsection{Sector Showers}\label{sec:SectorShowers}
In \cref{sec:GluonRadiation}, the use of the emitter-recoiler asymmetric mapping \cref{eq:radiationKinematics} was facilitated by partitioning the  antenna functions \cref{eq:Aqgq,eq:Aqgg,eq:Aggg} into two terms corresponding to the respective collinear sectors. An alternative to this approach would be to sectorise the phase space according to the two collinear regions, while leaving the eikonal intact. 
The sectorised versions of the branching kernels \cref{eq:Pqg,eq:Pgg} then read
\begin{align}
    P_{\q\g}^\mathrm{(sct)}(p_i,p_j,p_k) &= \frac{1}{s_{ij}}\left[\frac{2s_{ik}}{s_{jk}} + \frac{s_{jk}}{s_{ijk}}\right] \, , \label{eq:PqgSct} \\
    P_{\g\g}^\mathrm{(sct)}(p_i,p_j,p_k) &= \frac{1}{s_{ij}}\left[\frac{2s_{ik}}{s_{jk}} + \frac{2s_{jk}}{s_{ik}} + \frac{2s_{jk}s_{ik}}{s_{ijk}^2} \right] \, , \label{eq:PggSct}
\end{align}
and the $P_{\q\qbar}$ remains as in the ``global'' case. 
In particular, this avoids the dependence of the branching kernels on the global momentum $K$ and therefore on the azimuthal branching angle $\phi$.
Note that it would equally well be possible to add the corresponding purely collinear terms for the $jk$ sector to these functions.
The partitioning of the eikonal across the two antenna parents is then achieved by an additional sector veto,
\begin{equation}
    \Theta^\mathrm{(sct)}(p_i,p_j;n_j) = \theta\left(t_\mathrm{min} - t_{ij}\right) \quad \text{with} \quad t_\mathrm{min} = \min\limits_{i,j}\left\{ t_{ij}\right\} \, ,
\end{equation}
where the evolution variable is denoted as $t_{ij}\equiv t$ to highlight the dependence on the momenta $p_i$ and $p_j$. The minimum is taken across all radiator-emission pairs $(i,j)$.

By introducing the sector veto, the no-branching probability becomes
\begin{equation}
    \log\Delta_{n}^\mathrm{(sct)}(t,t') = -\sum\limits_{i}\sum\limits_{j\neq i}\, \int\limits^{t}_{t'} 8\uppi\alphas(t)\, \mc{C}_{ij}\, P_{ij}^{(\mathrm{sct})}(p_i,p_j,p_k)\, \Theta^{(\mathrm{sct})}(p_i,p_j;n_j)\,  \D\Phi^\mathrm{FF}_{ij} \, .
\end{equation}
Since sector showers always start from the minimal scale of the previous branching (without distortions by recoil effects) and both the radiation and splitting kinematics affect all momenta except $p_i$ and $p_j$ only by a Lorentz transformation, the post-branching sector resolution $t_\mathrm{min}$ can be overestimated by the shower restart scale.
After the trial branching only a veto against $t_{li}$ and $t_{jk}$, where $l$ and $k$ are the (leading-colour) partners of $i$ and $j$, respectively, must be implemented by means of the veto algorithm.
An implementation of sector showers in the \apollo framework will be left for future work.

\section{Matching to fixed-order calculations}\label{sec:FixedOrderCorrections}
Parton showers provide reliable predictions in softly and collinearly enhanced phase-space regions. Outside this limit, when final-state partons are well separated, fixed-order calculations provide accurate predictions. The two regimes can consistently be combined in matching and merging algorithms. In the following, a unitary approach to correct branching kernels to exact tree-level matrix elements is discussed in \cref{sec:MECs}. Its extension to a multiplicative NLO matching scheme will be discussed in \cref{sec:NLOMatching}.

\subsection{Matrix-element corrections}\label{sec:MECs}
For an infrared-safe observable $O$, the shower evolution is recursively defined in terms of a generating functional,
\begin{equation}
\begin{split}
    \mc{S}_{n}(t_n, O; \Phi_n) &= \Delta_n(t_n, t_\mathrm{c})O(\Phi_n) \\
    &\quad + \int\limits_{t_\mathrm{c}}^{t_n}\, \sum\limits_{i}\sum\limits_{j\neq i}\, 8\uppi\alphas(t)\, \mc{C}_{ij} P_{ij}(t,z,\phi;\Phi_n)\, \\
    &\quad\qquad \times \Delta_{n+1}(t_n,t)\, \mc{S}_{n+1}(t,O;\Phi_{n+1})\, \D \Phi^\mathrm{FF}_{ij}(t,z,\phi) \, ,
\end{split}
\label{eq:showerOperator}
\end{equation}
where $P_{ij}$ are the branching kernels in \cref{eq:Pqg,eq:Pgg,eq:Pqq} and $\D\Phi_{ij}^\mathrm{FF}$ denotes either the radiation or splitting phase-space measure. The sum runs over all leading-colour branchings where $\tilde{ij}$ is colour connected to $\tilde{k}$.
Here and in the following, the $n$-particle pre-branching (``Born'') configuration will be denoted by $\Phi_n \equiv \{\tilde{p}_1,\ldots,\tilde{p}_n\}$ and its corresponding tree-level squared matrix element as $B(\tilde{p}_1,\ldots,\tilde{p}_n)$.
The parton-shower prediction of the observable $O$ is then given in terms of the shower operator in \cref{eq:showerOperator} as
\begin{equation}
    \langle O\rangle^\mathrm{PS} = \int B(\tilde{p}_1,\ldots,\tilde{p}_n) O(\Phi_n)\, \mc{S}_n(t_n,O;\Phi_n)\, \D\Phi_n \,.
\end{equation}

Upon expanding the branching probability in \cref{eq:showerOperator} to first order in the strong coupling, it can be seen that the shower generates the $n+1$-parton configuration at scale $t$ with weight
\begin{equation}
    w_n^\mathrm{PS}(\Phi_{n+1}) = 8\uppi\alphas(t)\sum\limits_i\sum\limits_{j\neq i} \mc{C}_{ij}P_{ij}(t,z,\phi;\Phi_n) B(\tilde{p}_1,\ldots,\tilde{p}_n) \, ,
\end{equation}
which, in general, differs from the weight obtained from the full tree-level squared matrix element $R(p_1,\ldots,p_{n+1})$.
The correct LO weight is generated by introducing a MEC factor \cite{Bengtsson:1986et,Bengtsson:1986hr,Giele:2011cb},
\begin{equation}
    w_\mathrm{MEC} \equiv w_\mathrm{MEC}(p_1,\ldots,p_n) = \frac{R(p_1,\ldots,p_{n+1})}{\sum\limits_{i}\sum\limits_{j\neq i}\mc{C}_{ij}P_{ij}(p_i,p_j,p_k;K)B(\tilde{p}_1,\ldots,\tilde{p}_{n})} \, ,
\label{eq:MEC}
\end{equation}
meaning that the first shower branching is generated with no-branching probability
\begin{equation}
    \log \Delta^\mathrm{MEC}_{n}(t_n,t_{n+1}) = -\sum\limits_{i}\sum\limits_{j\neq i}\, \int\limits^{t_n}_{t_{n+1}} 8\uppi\alphas(t)\, w_\mathrm{MEC}\, \mc{C}_{ij}\, P_{ij}(p_i,p_j,p_k;K)\, \D\Phi^\mathrm{FF}_{ij}(t,z,\phi) \, .
\end{equation}
The correction factor \cref{eq:MEC} can be rewritten as an additive correction,
\begin{equation}
    w_\mathrm{MEC}(p_1,\ldots,p_n) = 1 + \frac{R(p_1,\ldots,p_{n+1}) - \sum\limits_{i}\sum\limits_{j\neq i}\mc{C}_{ij}P_{ij}(p_i,p_j,p_k;K)B(\tilde{p}_1,\ldots,\tilde{p}_{n})}{\sum\limits_{i}\sum\limits_{j\neq i}\mc{C}_{ij}P_{ij}(p_i,p_j,p_k;K)B(\tilde{p}_1,\ldots,\tilde{p}_{n})} \, .
\label{eq:MECadditive}
\end{equation}
The form of \cref{eq:MECadditive} suggests that the MEC weight becomes unity if the sum of shower histories reproduces the full matrix element. For simple processes, like first-order radiative corrections in colour-singlet two-body decays, this implies that the MEC can be absorbed into the definition of the branching kernels \cref{eq:Pqg,eq:Pgg,eq:Pqq}. In particular, this is facilitated by the use of partitioned antenna functions, which closely resemble physical matrix elements. 
Explicit expressions for matrix-element-corrected branching kernels implemented in the \apollo framework are collected in \cref{sec:MECBranchingKernels}.

For more complicated processes, especially those containing more than three coloured partons, the form of \cref{eq:MEC} still holds. However, for more than three coloured partons the squared matrix element contains subleading-colour contributions, while parton-shower algorithms usually work in the leading-colour limit and do not account for the additional subleading-colour singularity structure of the matrix element. 
Since subleading-colour corrections come with varying sign of the form $(-1/\NC^2)^\ell$, implementing subleading-colour corrections as separate branchings in the shower evolution would lead to negative event weights, which reduce the statistical significance of the event sample.
Subleading-colour corrections can be included in the shower evolution in a positive-definite way by realising that squared tree-level matrix elements are non-negative everywhere in phase-space and that the kinematics in \cref{eq:radiationKinematics,eq:splittingKinematics} depend explicitly only on the momenta $p_i^\mu$ and $p_j^\mu$.
It is therefore possible to define subleading-colour-corrected branching kernels as a sum over spectators $k$, which are now not necessarily colour-connected to $\tilde{ij}$ anymore,
\begin{equation}
    P^\mathrm{(cc)}_{ij}(p_i,p_j;K) = \sum\limits_{k\neq i,j} \mc{C}_{ij,k} P_{ij}(p_i,p_j,p_k;K) \, .
\label{eq:PijCC}
\end{equation}
Since both $t$ and $z$ are Lorentz-invariant quantities, their definition does not depend on the choice of reference frame. In particular, both $t$ and $z$ depend only on $p_i^\mu$, $p_j^\mu$, and a global momentum $K^\mu$. The colour-corrected branching kernels \cref{eq:PijCC} can therefore be overestimated by
\begin{equation}
    P^\mathrm{(cc)}_{ij}(p_i,p_j;K) \leq \sum\limits_{k\neq i,j} \mc{C}_{ij,k} P^\mathrm{trial}_{ij}(t,z) \, ,
\end{equation}
where $P^\mathrm{trial}_{ij}$ is given in \cref{eq:PtrialEmit} and \cref{eq:PtrialSplit} for gluon emissions and gluon splittings, respectively. This is positive by construction.
The colour-corrected branching kernels then act directly as trials for the full matrix element $R$ and the full-colour MEC weight becomes
\begin{equation}
    w_\mathrm{MEC}^{(\mathrm{FC})}(p_1,\ldots,p_n) = \frac{R(p_1,\ldots,p_{n+1})}{\sum\limits_{i}\sum\limits_{j\neq i}\sum\limits_{k\neq i,j}\mc{C}_{ij,k} P^\mathrm{trial}_{ij}(p_i,p_j;K)B(\tilde{p}_1,\ldots,\tilde{p}_{n})} \, .
\end{equation}
It is to be emphasised that the full-colour correction can only be applied for a finite number of branchings and is therefore inherently different to full-colour evolution approaches as e.g.~presented in \cite{Hoche:2020pxj,DeAngelis:2020rvq}.

\subsection{Multiplicative next-to-leading-order matching}\label{sec:NLOMatching}
The purpose of this section is to present two approaches to multiplicatively match the \apollo shower to NLO calculations, in a way that its logarithmic accuracy is retained. In particular, it is not the goal to analyse the resulting logarithmic structure of the matched shower beyond the statement that it acquires formal NLO accuracy for observables that are non-vanishing for the underlying ``Born'' configuration and that the matching does not interfere with the NLL structure of the shower.
It should be noted that additive NLO matching can straightforwardly be applied by taking into account the differences in the collinear evolution compared to \cite{Assi:2023rbu}.

Multiplicative matching approaches build upon the unitarity of a matrix-element corrected parton shower, cf.~\cite{Norrbin:2000uu,Nason:2004rx,Frixione:2007vw}. In particular, it is exploited that the first-order expansion of the matrix-element-corrected shower operator can be recast in the form of a projection-to-Born calculation\footnote{This omits contributions of the form $O(\Phi_{n+1})-O(\Phi_n)$ below the shower cutoff $t_\mathrm{c}$. These can be assumed to vanish for any infrared-safe observable $O$ when $t_\mathrm{c}$ is small.} \cite{Cacciari:2015jma},
\begin{align}
    \mc{S}_{n}^{\mathrm{MEC}}(t_n, O; \Phi_n) &= \left[1-\int\limits_{t_\mathrm{c}}^{t_n}\, \sum\limits_{i}\sum\limits_{j\neq i}\, 8\uppi\alphas(t_n)\, w_\mathrm{MEC}\, \mc{C}_{ij} P_{ij}(t,z,\phi;\Phi_n)\, \D \Phi^\mathrm{FF}_{ij}(t,z,\phi)\right]O(\Phi_n) \nonumber \\
    &\qquad + \int\limits_{t_\mathrm{c}}^{t_n}\, \sum\limits_{i}\sum\limits_{j\neq i}\, 8\uppi\alphas(t)\, w_\mathrm{MEC}\, \mc{C}_{ij} P_{ij}(t,z,\phi;\Phi_n)\, O(\Phi_{n+1})\, \D \Phi^\mathrm{FF}_{ij}(t,z,\phi) \nonumber \\
    &\qquad + \mathcal{O}(\alphas^2) \nonumber \\
    &= O(\Phi_n) + \int\limits_{t_\mathrm{c}}^{t_n}\, \sum\limits_{i}\sum\limits_{j\neq i}\, 8\uppi\alphas(t)\, w_\mathrm{MEC}\, \mc{C}_{ij} P_{ij}(t,z,\phi;\Phi_n) \label{eq:showerOperatorExp} \\
    &\hspace{3cm} \times \left[O(\Phi_{n+1}) - O(\Phi_n)\right]\, \D \Phi^\mathrm{FF}_{ij}(t,z,\phi) + \mathcal{O}(\alphas^2) \, . \nonumber
\end{align}
By reweighting Born-level phase-space points by a NLO weight fully differentially in the Born phase space, formal NLO accuracy is obtained for $n$-jet observables,
\begin{equation}
    \langle O\rangle^{\mathrm{NLO}+\mathrm{PS}} = \int K_\mathrm{NLO}(\tilde{p}_1,\ldots,\tilde{p}_n)B(\tilde{p}_1,\ldots,\tilde{p}_n) O(\Phi_n)\, \mc{S}_n(t_n,O;\Phi_n)\, \D\Phi_n \, .
\label{eq:NLOPSMaster}
\end{equation}
Upon expanding the shower operator as in \cref{eq:showerOperatorExp}, it can be seen that \cref{eq:NLOPSMaster} agrees with the NLO expression up to terms of $\mc{O}(\alphas^2)$.

Multiplicative matching schemes face two difficulties. On the one hand, the NLO Born weight $K_\mathrm{NLO}$ requires the integration of the real correction locally in a single Born phase-space point \cite{Frixione:2007vw,Platzer:2011bc,Hoche:2010pf}; on the other hand, a mismatch between the evolution leading to the first (hardest) branching and subsequent (shower) branchings is to be avoided \cite{Corke:2010zj,Hamilton:2020rcu}. In the following, two alternative schemes are presented, addressing both of these points. In \cref{sec:colourOrderNLOPS}, a refinement of the  dipole-shower schemes proposed in \cite{Frixione:2007vw,Hoche:2010pf} is presented, whereas in \cref{sec:bornLocalNLOPS}, a scheme more akin to the FKS scheme of \cite{Alioli:2010xd} is constructed in such a way as to avoid mismatches between the first and subsequent shower branchings.

\subsubsection{Matching with colour-ordered projectors}\label{sec:colourOrderNLOPS}
In the following, a real-radiation matrix element squared $R$ will be decomposed into individual colour layers $\ell$, each receiving contributions proportional to a single colour factor $\mc{C}^{(\ell)}$ only,
\begin{equation}
    R(p_1,\ldots,p_{n+1}) = \sum\limits_{\ell=0}^{m}\, \mc{C}^{(\ell)} R^{(\ell)}(p_1,\ldots,p_{n+1}) \, .
\label{eq:colourOrderedReal}
\end{equation}
In this context, $\ell=0$ denotes the leading-colour layer, $\ell=1$ the subleading-colour layer, $\ell=2$ the subsubleading-colour layer, and so on. In each colour layer, the terms $R^{(\ell)}$ may receive contributions from different ``orderings'' of the momenta $\{p_1,\ldots,p_{n+1}\}$. 
A Born-local NLO weight can then be calculated by introducing a decomposition of unity for each of the colour layers separately,
\begin{equation}
    1 = \sum\limits_{i}\sum\limits_{j\neq 0} W^{(\ell)}_{ij} \, ,
\end{equation}
where the sum runs over all pairs $(i,j)$ that correspond to a valid branching $\tilde{ij}\mapsto i+j$ in this colour layer. In the following, the product of the shower branching kernel with the reduced matrix element is denoted as $K_{ij}$,
\begin{equation}
\begin{split}
    K_{ij}^{(\ell)}(p_1,\ldots,p_{n+1}) &= 8\uppi\alphas\, \frac{1}{s_{ij}}\left[\frac{2s_{ik}(p_in_j)}{s_{jk}(p_in_j)+s_{ij}(p_kn_j)} - \frac{2z}{1-z}\right]B(\tilde{p}_1,\ldots,\tilde{p}_n) \\
    &\qquad + 8\uppi\alphas\, P_{ij}^{\mu\nu}(p_i,p_j,p_k;K)B_{\mu\nu}(\tilde{p}_1,\ldots,\tilde{p}_{n}) \, .
\end{split}
\end{equation}
where $z$ is defined as in \cref{eq:zDefEmit} and $P^{\mu\nu}_{ij}$ denotes spin-dependent versions of the shower branching kernels \cref{eq:Pqg,eq:Pgg,eq:Pqq}\footnote{Note that for simplicity, the $q\mapsto qg$ branching kernel has been included in this notation.}.
Assuming that all $K_{ij}$ are positive definite, which is true for all branching kernels \cref{eq:Pqg,eq:Pgg,eq:Pqq}, the matrix elements $R^{(\ell)}$ can be decomposed as
\begin{equation}
    R^{(\ell)}(p_1,\ldots,p_{n+1}) = \sum\limits_{i}\sum\limits_{j\neq i} W^{(\ell)}_{ij} R^{(\ell)}(p_1,\ldots,p_{n+1}) \, ,
\label{eq:colourOrderedRealDecomposition}
\end{equation}
where
\begin{equation}
    W^{(\ell)}_{ij} = \frac{K^{(\ell)}_{ij}(p_1,\ldots,p_{n+1})}{\sum\limits_i\sum\limits_{j\neq i} K^{(\ell)}_{ij}(p_1,\ldots,p_{n+1})} \,.
\label{eq:apolloProjectors}
\end{equation}
This is only well-defined if the sum in the denominator is guaranteed to be positive definite. As mentioned above, this is true for all shower branching kernels defined in \cref{eq:Pqg,eq:Pgg,eq:Pqq} by construction. Due to the separate construction of the weights $W^{(\ell)}_{ij}$ for each colour layer, all of the terms $K^{(\ell)}_{ij}$ contribute with the same colour factor, i.e., there are no variations in sign in the sum in the denominator of \cref{eq:apolloProjectors}. \Cref{eq:colourOrderedRealDecomposition} is therefore well-defined as long as the Born matrix elements squared $B$ are positive definite.

The integral of the branching kernels, which is needed to cancel explicit singularities in the virtual correction $V$, can straightforwardly be computed as
\begin{equation}
    \mc{K}^{(\ell)}_{ij}(\tilde{p}_1,\ldots,\tilde{p}_n) = \int\, P_{ij}(p_i,p_j,p_k;K)\, \D\Phi^\mathrm{FF}_{ij}(p_i,p_j;K;\varepsilon)\, B(\tilde{p}_{1},\ldots,\tilde{p}_n) \, ,
\end{equation}
where $\D\Phi^\mathrm{FF}_{ij}(p_i,p_j;K;\varepsilon)$ denotes the $D=4-2\varepsilon$ dimensional phase-space measure in the emission and splitting kinematics for gluon emissions and gluon splittings, respectively. The integrals of the gluon-splitting branching kernels with splitting kinematics as well as the integrals over the soft parts of the branching kernels \cref{eq:Pqg,eq:Pgg} with radiation kinematics have been derived in \cite{Assi:2023rbu}. The remaining integrals over the purely collinear pieces are left for future work.

In terms of the colour-ordered projectors \cref{eq:apolloProjectors}, the $\bar{B}$ function can now be written as,
\begin{equation}
\begin{split}
    \bar{B}(p_1,\ldots,p_n) &= B(p_1,\ldots,p_n) + V(p_1,\ldots,p_n) + \sum\limits_{\ell=0}^m\mc{C}^{(\ell)}\sum\limits_i\sum\limits_{j\neq i} \mc{K}^{(\ell)}_{ij}(p_1,\ldots,p_n) \\
    &\quad + \sum\limits_{i}\sum\limits_{j\neq i}\, \sum\limits_{\ell=0}^{m}\,\mc{C}^{(\ell)} \int \D\Phi^\mathrm{FF}_{ij}\, \left[W^{(\ell)}_{ij}R^{(\ell)}(p_1,\ldots,p_{n+1}) - K^{(\ell)}_{ij}(p_1,\ldots,p_{n+1})\right] \, .
\end{split}
\label{eq:weightNLOColourOrderedProjectors}
\end{equation}
Note that in the real correction, the sum over colour layers $\ell$ can be taken within each radiator-emission pair $ij$, as the emission kinematics in \cref{eq:radiationKinematics} and the splitting kinematics in \cref{eq:splittingKinematics} only depend on $i$ and $j$ and the global momentum $K$. 

Finally, it should be emphasised that the explicit decomposition into colour layers does not pose a challenge for an automated implementation, as most automated matrix-element generators anyway calculate QCD processes using the colour-flow picture, which straightforwardly allows for the extraction of the necessary colour factors and associated squared amplitudes.

\begin{figure}[t]
    \centering
    \includegraphics[width=0.45\textwidth]{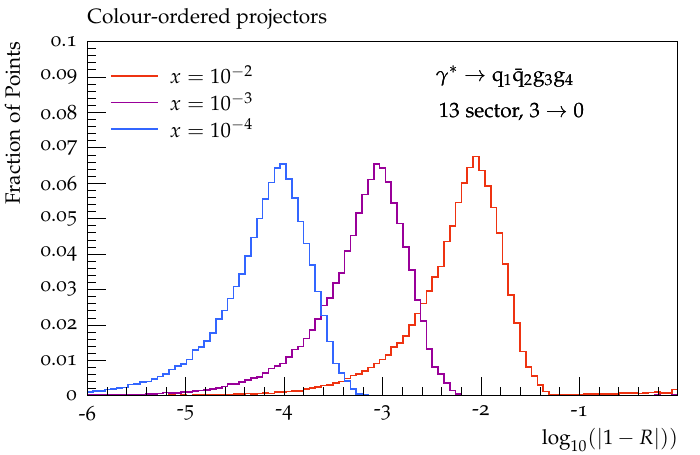}
    \includegraphics[width=0.45\textwidth]{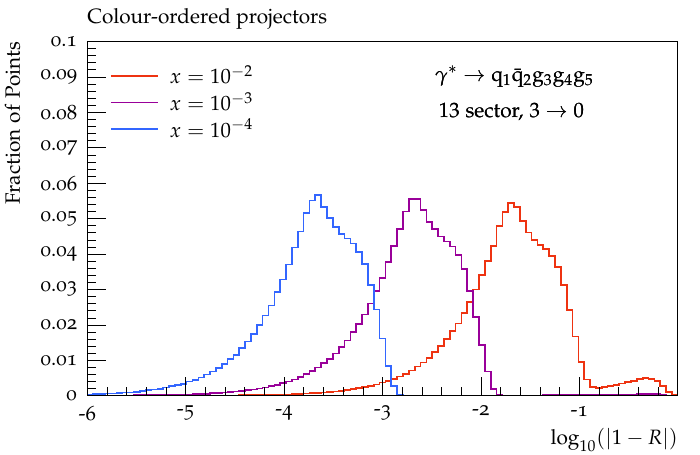}\\
    \includegraphics[width=0.45\textwidth]{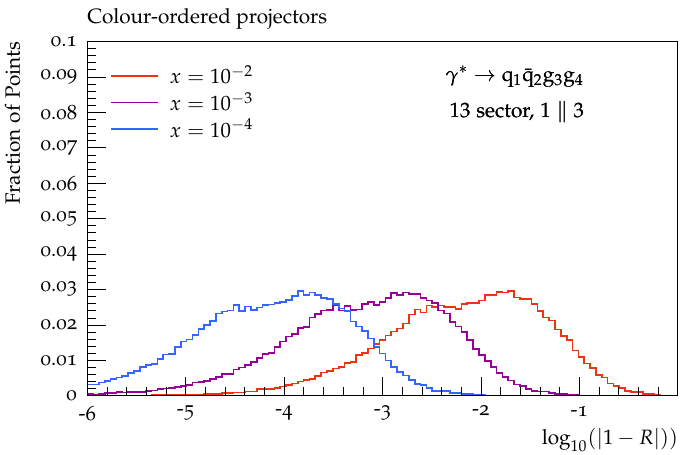}
    \includegraphics[width=0.45\textwidth]{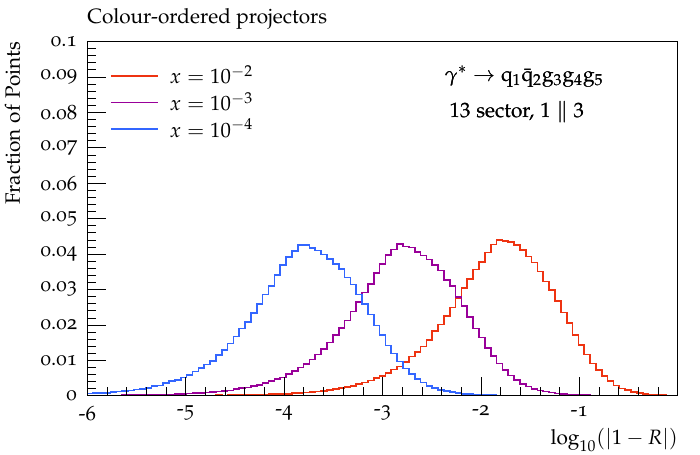}
    \caption{Representative examples of the numerical convergence of the subtraction term against the projected matrix element in the soft-gluon (top) and quark-gluon collinear (bottom) limits.} 
    \label{fig:numConvPS}
\end{figure}

To close the discussion of multiplicative matching with colour-ordered projectors, the feasibility of the construction of viable local subtraction terms shall be exemplified. To this end, the numerical convergence of the second line of \cref{eq:weightNLOColourOrderedProjectors} is tested in single-unresolved limits for $\upgamma^*\to \mathrm{q}\bar{\mathrm{q}}\mathrm{g}\mathrm{g}$ and $\upgamma^*\to \mathrm{q}\bar{\mathrm{q}}\mathrm{g}\mathrm{g}\mathrm{g}$, which contribute to the real correction in the processes $\upgamma^* \to 3j$ and $\upgamma^*\to 4j$, respectively.
This is done by testing the numerical agreement between the projected matrix element and the subtraction term in trajectories towards the single-unresolved limit. These are obtained by generating configurations with $n+1$ resolved partons, where $n$ is the number of partons in the Born-level process, and rescaling the energy as $E_j\to xE_j$ or the invariant mass $\sqrt{s_{ij}}\to x\sqrt{s_{ij}}$ in the $j$-soft or $i\parallel j$ limit, respectively.
The convergence is expressed in terms of the number of agreeing digits, calculated as
\begin{equation}
    \log_{10}(\vert 1-R\vert) \, , \quad \text{where} \quad R = \frac{\sum\limits_{\ell=0}^mK_{ij}^{(\ell)}(p_1,\ldots,p_{n+1})}{\sum\limits_{\ell=0}^m W_{ij}^{(\ell)}R^{(\ell)}(p_1,\ldots,p_{n+1})} \, .
\end{equation}
Representative examples for these tests can be found in \cref{fig:numConvPS}, where the $\q\parallel\g$ and soft-gluon limit of a single phase-space sector is shown. Both limits receive contributions from all colour layers and therefore highlight the applicability of the projectors to processes with a non-trivial colour structure. In all cases, the figures show very good convergence of the subtraction term against the projected matrix element with decreasing $x$.

\subsubsection{Matching with Born-local subtraction terms}\label{sec:bornLocalNLOPS}
The multiplicative matching discussed in the previous subsection requires detailed knowledge of the colour structure and behaviour of the squared matrix elements entering the calculation.
In the following, an alternative matching scheme is constructed, in close analogy to the local analytic sector subtraction at NLO, cf.~\cite{Magnea:2018hab}.
To this end, some notation of \cite{Braun-White:2023sgd} and \cite{Magnea:2018hab} is adopted. Specifically, a projector into the soft limit of particle $j$ is denoted as $\boldsymbol{S}^\downarrow_j$ and a projector into the collinear limit of particles $i$ and $j$ is denoted as $\boldsymbol{C}^\downarrow_{ij}$. 
By introducing a decomposition of unity,
\begin{equation}
    1 = \sum\limits_{i=1}^{n+1}\sum\limits_{j\neq i}^{n+1} \Wsct_{ij} \label{eq:sectorRequirement1}
\end{equation}
one can then proceed and decompose the real matrix element $R(p_1,\ldots,p_{n+1})$ into a sum over sectors,
\begin{equation}
    R(p_1,\ldots,p_{n+1}) = \sum\limits_{i=1}^{n+1}\sum\limits_{j\neq i}^{n+1} \Wsct_{ij}R(p_1,\ldots,p_{n+1}) \, ,
\end{equation}
In analogy to the FKS subtraction scheme \cite{Frixione:2002ik}, each of the sector functions $\Wsct_{ij}$ should single out at most one collinear and one soft limit,
\begin{equation}
    \boldsymbol{S}^\downarrow_{j}\Wsct_{kl} = \delta_{jl} \, , \quad \boldsymbol{C}^\downarrow_{ij}\Wsct_{kl} = \delta_{ik}\delta_{jl} + \delta_{il}\delta_{jk} \, . \label{eq:sectorRequirement2}
\end{equation}
In order to simplify the integration of the subtraction terms, the sector functions $\Wsct_{ij}$ are subject to the following requirements,
\begin{equation}
    \boldsymbol{S}^\downarrow_{j}\sum\limits_{i=1}^{n+1}\Wsct_{ij} = 1 \, , \quad \boldsymbol{C}^\downarrow_{ij}\left(\Wsct_{ij}+\Wsct_{ji}\right) = 1 \, . \label{eq:sectorRequirement3}
\end{equation}
As long as they obey the requirements \cref{eq:sectorRequirement1,eq:sectorRequirement2,eq:sectorRequirement3}, the definition of the sector functions $\Wsct_{ij}$ is arbitrary. In particular, they can either be smooth functions or step functions. A particularly simple choice is given by \cite{Magnea:2018hab}
\begin{equation}
    \Wsct_{ij} = \frac{(2p_iK)(2p_jK)}{E_js_{ij}s} = \frac{2}{E_j(1-\cos\theta_{ij})}\, ,
\end{equation}
where the right-hand side is taken in the event centre-of-mass frame.

Subtraction terms are then constructed as \cite{Magnea:2018hab}
\begin{equation}
    K_{ij}(p_1,\ldots,p_{n+1}) = \left(\boldsymbol{S}^\downarrow_{j} + \boldsymbol{C}^\downarrow_{ij} - \boldsymbol{S}^\downarrow_{j}\boldsymbol{C}^\downarrow_{ij}\right) R(p_1,\ldots,p_{n+1})\Wsct_{ij} \, .
\label{eq:bornLocalSubtractionTerm}
\end{equation}
The definition of the subtraction term in \cref{eq:bornLocalSubtractionTerm} is subject to the choice of a suitable momentum mapping $\{p_1,\ldots,p_i, p_j,\ldots,p_{n+1}\} \to \{\tilde{p}_{ij}, \ldots, \tilde{p}_{ij},\ldots, \tilde{p}_{n}\}$. Ideally, the same kinematics as in the shower evolution are employed in the generation of the Born-local NLO weight. However, the gluon-emission kinematics of \cref{eq:radiationKinematics} are intrinsically asymmetric in $i$ and $j$, so that $ij$ and $ji$ sectors would lead to different reduced (Born) configurations. In particular, this implies that the corresponding subtraction terms would need to be integrated separately over the two sectors, including an undesirable dependence on the respective sector functions $W_{ij}$ and $W_{ji}$. 
In the following, a different strategy is adopted. Since the splitting kinematics in \cref{eq:splittingKinematics} are symmetric in $i$ and $j$ by construction, it provides a natural choice for the construction of Born-local subtraction terms.

The inverse of the splitting kinematics in \cref{eq:splittingKinematics} are constructed in \cref{app:kinematics}.
In terms of \cref{eq:splittingKinematicsInverse}, the individual contributions to \cref{eq:bornLocalSubtractionTerm} are given by
\begin{align}
    \boldsymbol{S}^\downarrow_{j}R(p_1,\ldots,p_{n+1}) &= -8\uppi\alphas\, \sum\limits_{i\neq j}^{n+1}\sum\limits_{k\neq j}^{n+1} \Big\langle \tilde{p}_1, \ldots, \tilde{p}_{n} \Big\vert \mathbf{T}_i\mathbf{T}_k \frac{2s_{ik}}{s_{ij}s_{jk}} \Big\vert \tilde{p}_1,\ldots,\tilde{p}_{n} \Big\rangle \, , \\
    \boldsymbol{C}^\downarrow_{ij}R(p_1,\ldots,p_{n+1}) &= 8\uppi\alphas\, \sum\limits_{\lambda,\lambda' = \pm} \Big\langle \tilde{p}_1, \ldots, \tilde{p}_{n} \Big\vert \mathbf{T}_{\widetilde{ij}}^2 \frac{P_{ij}^{\lambda\lambda'}(\bar{z})}{s_{ij}} \Big\vert \tilde{p}_1,\ldots,\tilde{p}_{n} \Big\rangle\, , \\
    \boldsymbol{S}^\downarrow_{j}\boldsymbol{C}^\downarrow_{ij}R(p_1,\ldots,p_{n+1}) &= 8\uppi\alphas\, \Big\langle \tilde{p}_1, \ldots, \tilde{p}_{n} \Big\vert \mathbf{T}_{\widetilde{ij}}^2 \frac{1}{s_{ij}}\frac{2\bar{z}}{1-\bar{z}} \Big\vert \tilde{p}_1,\ldots,\tilde{p}_{n} \Big\rangle\, ,
\end{align}
with the spin-dependent DGLAP splitting functions $P_{ij}^{\lambda\lambda'}(\bar{z})$ and $\bar{z}$ as in \cref{eq:splittingKinVariables}, i.e.,
\begin{equation}
    \bar{z} = \frac{p_i K}{(p_i+p_j)K} \, .
\end{equation}
Finally, the subtraction terms \cref{eq:bornLocalSubtractionTerm} have to be integrated over the $D=4-2\varepsilon$ dimensional splitting phase space, so as to cancel explicit poles in the virtual correction $V$. By summing over all sectors $ij$, the subtraction term \cref{eq:bornLocalSubtractionTerm} can be brought into a form in which all sector functions have been summed, cf.~\cite{Magnea:2018hab},
\begin{equation}
\begin{split}
    K(p_1,\ldots,p_{n+1}) &= \sum\limits_{i=1}^{n+1}\sum\limits_{j\neq i}^{n+1} K_{ij}(p_1,\ldots,p_{n+1}) \\
    &= \sum\limits_{j=1}^{n+1}\boldsymbol{S}^\downarrow_{j}R(p_1,\ldots,p_{n+1}) \\
    &\quad + \sum\limits_{i=1}^{n+1}\sum\limits_{j>i}^{n+1} \boldsymbol{C}^\downarrow_{ij}\left(1 - \boldsymbol{S}^\downarrow_{i} - \boldsymbol{S}^\downarrow_{j}\right) R(p_1,\ldots,p_{n+1}) \, .
\end{split}
\label{eq:bornLocalSubtractionTermFull}
\end{equation}
This implies that the soft and purely collinear pieces can be integrated separately over the entire kinematical range without having to account for the presence of sector functions,
\begin{equation}
\begin{split}
    \mc{K}(\tilde{p}_1,\ldots,\tilde{p}_{n}) &= -8\uppi\alphas\, \sum\limits_{j=1}^{n+1}\sum\limits_{i\neq j}^{n+1}\sum\limits_{k\neq j}^{n+1} \int \frac{2s_{ik}}{s_{ij}s_{jk}} \D\Phi_\mathrm{split}^\mathrm{FF}(p_i,p_j;K,\varepsilon)\, \Big\langle \tilde{p}_1, \ldots, \tilde{p}_{n} \Big\vert \mathbf{T}_i\mathbf{T}_k \Big\vert \tilde{p}_1,\ldots,\tilde{p}_{n} \Big\rangle\,  \\
    &\quad + 8\uppi\alphas\, \sum\limits_{i=1}^{n+1}\sum\limits_{j\neq i}^{n+1} \int \frac{1}{s_{ij}}P_{ij}^\mathrm{(coll)}(\bar{z},\varepsilon)\, \D\Phi_\mathrm{split}^\mathrm{FF}(p_i,p_j;K,\varepsilon)\, B(\tilde{p}_1,\ldots,\tilde{p}_n) \, ,
\end{split}
\end{equation}
where $P_{ij}^{(\mathrm{coll})}$ denotes the purely collinear part of the splitting functions,
\begin{align}
    P_{qg}^{(\mathrm{coll})}(\bar{z},\varepsilon) &= (1-\varepsilon)(1-\bar{z}) \, ,\\
    P_{gg}^{(\mathrm{coll})}(\bar{z},\varepsilon) &= 2\bar{z}(1-\bar{z}) \, ,\\
    P_{q\bar{q}}^{(\mathrm{coll})}(\bar{z},\varepsilon) &= 1-\frac{2\bar{z}(1-\bar{z})}{1-\varepsilon} \, .
\end{align}
The integrals of the purely collinear parts of the splitting functions with splitting kinematics have been derived in \cite{Assi:2023rbu}. The corresponding integrals of the soft eikonal will be derived in future work.
It is to be highlighted that the summation in \cref{eq:bornLocalSubtractionTermFull} is facilitated by the explicit symmetry of the splitting map \cref{eq:splittingKinematicsInverse}.

Using the subtraction term in \cref{eq:bornLocalSubtractionTerm}, NLO-weighted Born phase-space points can be generated according to
\begin{equation}
\begin{split}
    \bar{B}(\tilde{p}_1,\ldots,\tilde{p}_n) &= B(\tilde{p}_1,\ldots,\tilde{p}_n) + V(\tilde{p}_1,\ldots,\tilde{p}_n) + \sum\limits_j \mc{K}_{ij}(\tilde{p}_1,\ldots,\tilde{p}_n) \\
    &\quad + \sum\limits_{i=1}^{n+1}\sum\limits_{j\neq i}^{n+1} \int \D \Phi_\mathrm{split}^\mathrm{FF}\,\Bigg[R(p_1,\ldots,p_{n+1})\Wsct_{ij} - K_{ij}(p_1,\ldots,p_{n+1})\Bigg] \, .
\end{split}
\label{eq:weightNLOBornLocal}
\end{equation}
That is, by first generating the momenta $\{\tilde{p}_1,\ldots,\tilde{p}_n\}$ and then looping over all possible sectors in the real phase space that lead to the exact same Born-level phase-space point.

As alluded to in \cref{sec:AntennaShowers}, the splitting kinematics in \cref{eq:splittingKinematics} are not NLL safe as a result of the local assignment of the transverse recoil. In the following it will be argued how the matching outlined in this section anyway retains the logarithmic accuracy of the shower.
By construction, the first shower branching is corrected to the exact real-radiation matrix element everywhere in phase space, cf.~\cref{eq:MEC}. In particular, this means that its expansion reproduces the correct tree-level matrix element after the first branching. In order to not spoil the logarithmic accuracy of the shower, the strongly-ordered tree-level matrix element must also be reproduced upon subsequent branchings for all configurations where branchings are ordered in angle but have a commensurate evolution variable \cite{Dasgupta:2020fwr}. This is achieved by generating subsequent gluon emissions using the  kinematics in \cref{eq:radiationKinematics}, which have been demonstrated in \cite{Herren:2022jej} to obey this requirement. At this point it should be stressed that the first branching (including the MEC) and all subsequent branchings are generated using the exact same definition of the evolution variable, namely 
\begin{equation}
    t=(1-z)s_{ij} \, ,
\end{equation}
cf.~\cref{eq:showerEvolEmit,eq:showerEvolSplit}, and using the same branching kernels \cref{eq:Pqg,eq:Pgg,eq:Pqq}. In particular, this implies that there exists no mismatch between the ``hardest emission generator'' and the parton-shower algorithm, an important aspect to retain the logarithmic accuracy of the shower and to avoid the use of vetoed showers \cite{Corke:2010zj,Hamilton:2023dwb}. The first and subsequent branchings merely differ in the assignment of the transverse recoil, which is corrected by the MEC in case of the former.

\begin{figure}[t]
    \centering
    \includegraphics[width=0.45\textwidth]{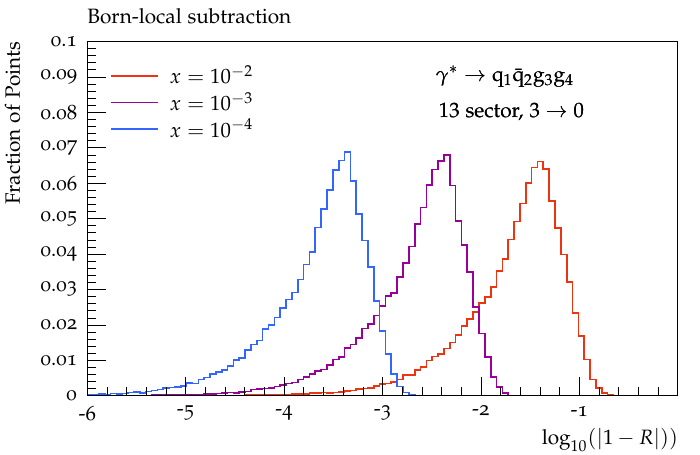}
    \includegraphics[width=0.45\textwidth]{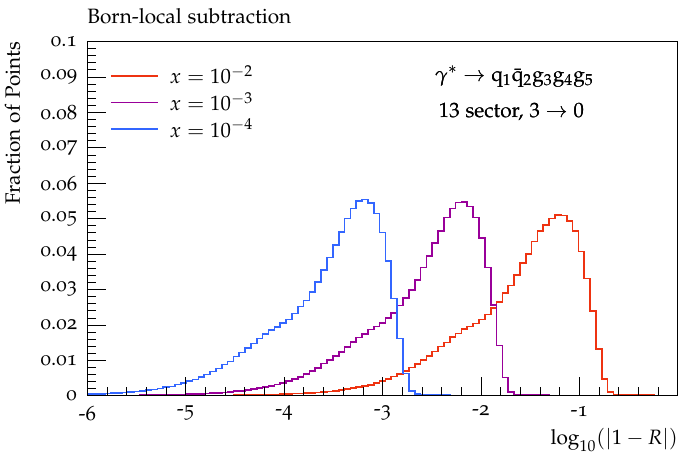}\\
    \includegraphics[width=0.45\textwidth]{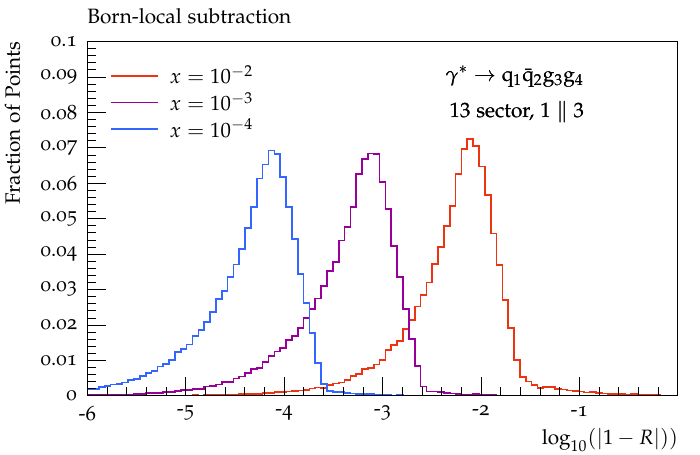}
    \includegraphics[width=0.45\textwidth]{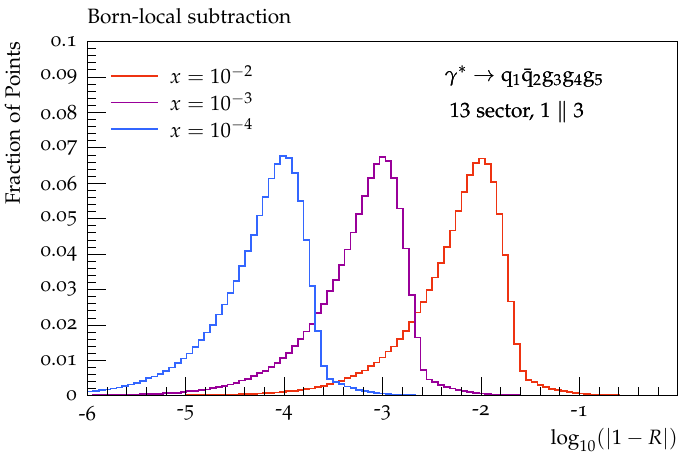}
    \caption{Representative examples of the numerical convergence of the subtraction term against the projected matrix element in the soft-gluon (top) and quark-gluon collinear (bottom) limits.} 
    \label{fig:numConvSct}
\end{figure}

This section shall be closed by highlighting the feasibility of the construction of viable Born-local subtraction terms according to \cref{eq:bornLocalSubtractionTerm} for non-trivial processes.
This is done by testing the numerical convergence of the second line of \cref{eq:weightNLOBornLocal} for the two processes $\upgamma^*\to \mathrm{q}\bar{\mathrm{q}}\mathrm{g}\mathrm{g}$ and $\upgamma^*\to \mathrm{q}\bar{\mathrm{q}}\mathrm{g}\mathrm{g}\mathrm{g}$, constituting parts of the real correction to $\upgamma^*\to 3j$ and $\upgamma^*\to 4j$, respectively. 
To this end, trajectories into the single-unresolved limits are constructed by uniformly generating phase-space points with $n+1$ resolved partons, where $n$ denotes the number of partons in the Born-level process, and rescaling either the energy $E_j \to xE_j$ in the $j$-soft limit or the invariant mass $\sqrt{s_{ij}}\to x\sqrt{s_{ij}}$ in the $i\parallel j$ limit. The convergence is assessed by calculating the number of digits to which the (projected) matrix element and the subtraction term agree as
\begin{equation}
    \log_{10}(\vert 1-R\vert) \, , \quad \text{where} \quad R = \frac{K_{ij}(p_1,\ldots,p_{n+1})}{W_{ij}^{\mathrm{sct}}R(p_1,\ldots,p_{n+1})} \, .
\end{equation}
\Cref{fig:numConvSct} contains representative examples for both processes, showing the convergence in the $q\parallel g$ and soft-gluon limit. These limits receive contributions from all colour layers. In all cases, the convergence of the subtraction term against the projected matrix element is clearly visible with decreasing $x$.

\section{Validation and preliminary results}\label{sec:ValidationResults}
In this section, the consistency of the \apollo shower with NLL resummation for a wide range of global event-shape observables is explicitly validated through numerical tests in \cref{sec:NumericalValidation}. The implementation in the \pythia event generator is checked in \cref{sec:ComparisonPythia}, which contrasts the \apollo algorithm with \pythia's existing ``simple'' shower and the \vincia antenna shower.

\subsection{Numerical validation of the logarithmic accuracy}\label{sec:NumericalValidation}
The logarithmic structure pertaining to the global recoil scheme in \cref{eq:radiationKinematics} has been analysed analytically in great detail in \cite{Herren:2022jej}. As the \apollo shower does not differ from \alaric in the way it assigns the transverse momentum in gluon emissions, this exercise is not repeated here. It should be highlighted that the \apollo shower differs from the \alaric algorithm described in \cite{Herren:2022jej,Assi:2023rbu} in the evolution variable and the treatment of purely collinear terms in the branching kernels. 

It is instructive to explicitly verify that the shower algorithm described in \cref{sec:AntennaShowers} is indeed consistent with NLL resummation. To this end, the approach of \cite{Dasgupta:2020fwr} is followed, i.e., the shower algorithm is contrasted to the NLL result in the limit $\alphas \to 0$ with fixed $\alphas\log v$, where $v$ denotes the value of an observable $V$ that can be parametrised as \cite{Banfi:2004yd}
\begin{equation}
    V(k) = \left(\frac{k_\mathrm{t}}{\sqrt{s}}\right)^a \mathrm{e}^{-b \eta_k} \, .
\end{equation}
Numerically, this is achieved by implementing both the \apollo and \vincia final-state shower algorithms in the \texttt{PyPy} framework developed within the scope of \cite{Herren:2022jej}. For the sake of numerical stability, only the particular choice of $c=s_{jk}/(s_{ij}+s_{jk})$ is implemented, which differs from the ``Kosower'' rotation considered in the \pythia implementation of \vincia, cf.~\cite{Giele:2007di,Brooks:2020upa}. This allows to construct the post-branching momenta directly in the event frame and avoids boosts into the centre-of-momentum frame of antenna ends with vanishing invariant mass.
The limit $\alphas \to 0$ is extracted by a bin-wise linear regression fit of the results obtained for $\alphas \in \{0.02, 0.01, 0.005, 0.0025\}$.
As already highlighted in \cite{Dasgupta:2020fwr,Herren:2022jej}, it suffices to consider the strict leading-colour limit $2\CF=\CA=3$ and keep the value of the strong coupling $\alphas$ fixed to reproduce the failure of local transverse-recoil schemes \cref{eq:antennaKinematics} and establish the consistency of the global transverse-recoil scheme \cref{eq:radiationKinematics} with the NLL target.
In the following, these simplifications are therefore applied to obtain both the NLL reference results and the parton-shower predictions.

\begin{figure}[t]
    \centering
    \includegraphics[width=0.6\textwidth]{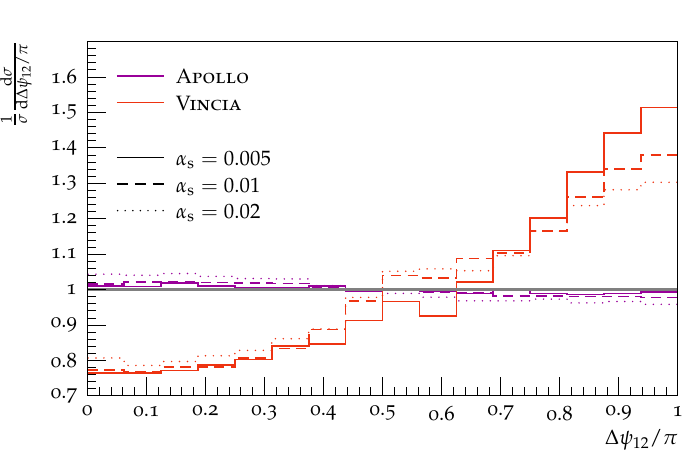}
    \caption{Numerical predictions in the limit of small $\alphas$ for the angle $\Delta\psi_{12}$ normalised to the NLL result obtained with \vincia and \apollo.}
    \label{fig:validationDPsi}
\end{figure}

The inconsistency of the antenna recoil scheme with NLL resummation is best seen in the distribution of the angular separation of the two primary Lund-plane declusterings, $\Delta\psi_{12}$ \cite{Dasgupta:2020fwr}. \Cref{fig:validationDPsi} contains predictions obtained with the conventional antenna shower \vincia and the new partitioned dipole-antenna shower \apollo, both normalised to the NLL result. The \vincia results in \cref{fig:validationDPsi} qualitatively reproduce the features observed in both \cite{Dasgupta:2020fwr} and \cite{Herren:2022jej} for the case of the \pythia and \dire showers. In particular, there remains a notable angular dependence in the limit $\alphas\to 0$, in disagreement with the flat NLL result. The flat $\Delta\psi_{12}$ distribution is, on the other hand, reasonably well reproduced by the \apollo shower for all values of $\alphas$ considered here.
Since this test is sufficient to highlight the shortcomings of the antenna recoil scheme, predictions obtained with \vincia will not be shown for the other event shapes.

\begin{figure}[t]
    \centering
    \includegraphics[width=0.45\textwidth]{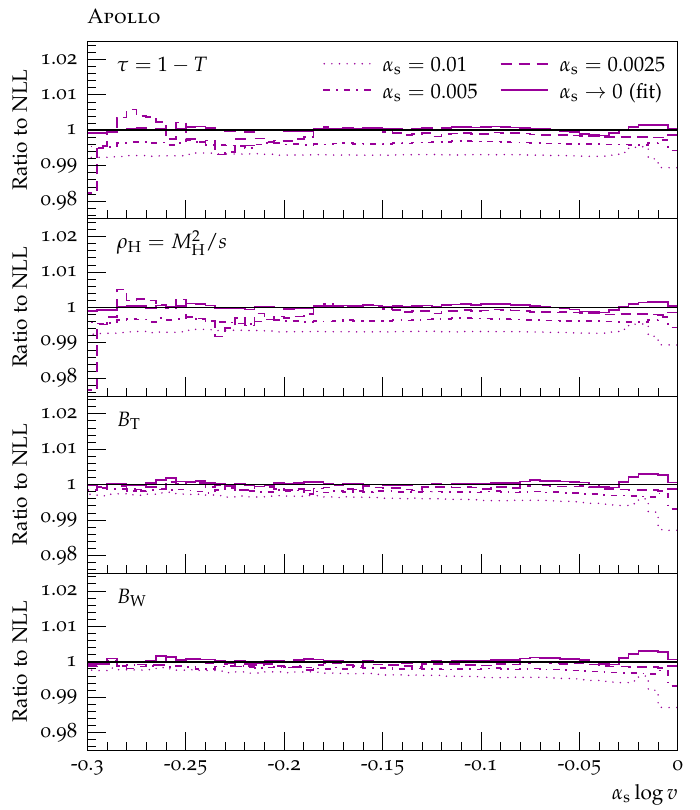}
    \includegraphics[width=0.45\textwidth]{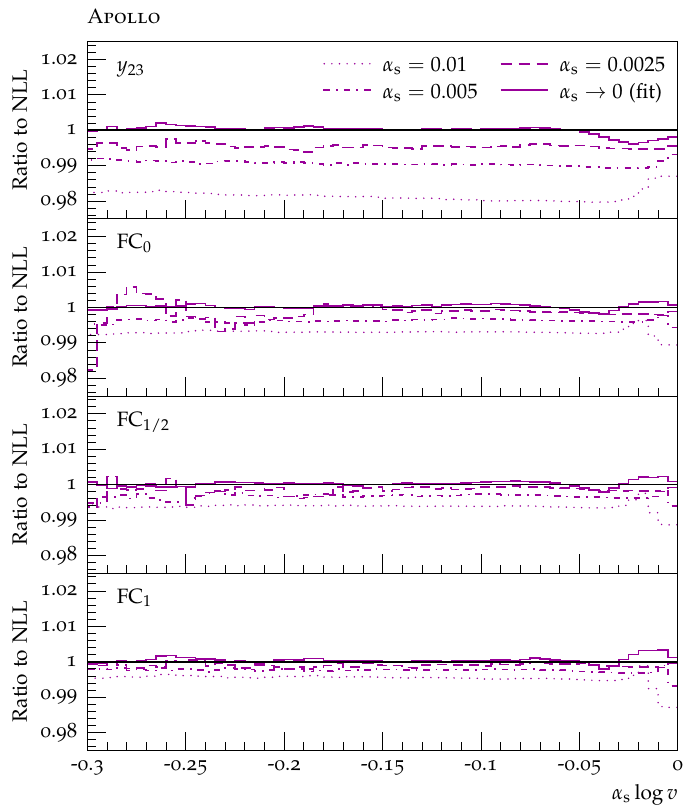}
    \caption{\apollo predictions in the limit of small $\alphas$ for event shapes normalised to the NLL result.}
    \label{fig:validationNLL}
\end{figure}

\Cref{fig:validationNLL} contains predictions obtained with the \apollo algorithm for the classical event-shapes (one minus) thrust $\tau$ \cite{Brandt:1964sa,Farhi:1977sg}, heavy-jet mass $\rho_\mathrm{H}$ \cite{Clavelli:1979md}, total and wide jet broadening $B_\mathrm{T}$ and $B_\mathrm{W}$ \cite{Rakow:1981qn,Catani:1992jc}, and the three-jet resolution $y_{23}$ in the Cambridge algorithm \cite{Dokshitzer:1997in}. In addition, also the fractional energy correlators $\mathrm{FC}_x$ \cite{Banfi:2004yd} are considered for the three choices $0$, $\frac{1}{2}$, and $1$ of the parameter $x$.
In all cases, the difference to the NLL reference result decreases with decreasing $\alphas$. Disregarding remaining statistical fluctuations and a residual bump at large $L$ in the $\alphas\to 0$ fit, owing to the regression across the $\alphas = 0.02$ and $\alphas = 0.01$ results, a constant ratio of $1$ is found in the $\alphas \to 0$ limit in all cases.
This highlights consistency of the \apollo algorithm with NLL resummation.

\subsection{Preliminary results}\label{sec:ComparisonPythia}
In this section, the \apollo shower is compared to LEP data, contrasting it to existing showers in \pythia. Specifically, predictions obtained with \pythia's default ``simple shower'', referred to as \pythia in the following, and the \vincia antenna shower are shown alongside predictions obtained with \apollo.

Predictions are shown at hadron level, employing the current default tunes for \pythia and \vincia. For \pythia, this corresponds to the Monash 2013 tune \cite{Skands:2014pea} with a one-loop running strong coupling with a value of $\alphas(m_\mathrm{Z})=0.1365$ in the $\overline{\mathrm{MS}}$ scheme and a shower cutoff of $t_\mathrm{c} = (0.5~\mathrm{G}e\mathrm{V})^2$.
In the case of \vincia, the default tune \cite{Brooks:2020upa} is used with shower cutoff $t_\mathrm{c} = (0.75~\mathrm{G}e\mathrm{V})^2$ and a two-loop running for the strong coupling with a value of $\alphas(m_\mathrm{Z})=0.118$ in the $\overline{\mathrm{MS}}$ scheme. The effective value of the coupling is calculated in the CMW scheme \cite{Catani:1990rr}, including additional renormalisation-scale factors for gluon emissions, $k_\mathrm{emit} = 0.66$, and gluon splittings, $k_\mathrm{split}=0.8$.

\apollo predictions are obtained with a shower cutoff of $t_\mathrm{c} = (0.75~\mathrm{G}e\mathrm{V})^2$ and a value of $\alphas(m_\mathrm{Z})=0.118$ in the $\overline{\mathrm{MS}}$ scheme. The effective coupling is obtained in the CMW scheme with two-loop running.
The input parameters for the string-fragmentation model are chosen as in the default \vincia tune, except $\texttt{StringZ:aLund = 0.44}$ and $\texttt{StringZ:bLund = 0.55}$.
Lacking a dedicated tune of the \apollo shower, the uncertainty associated to the choice of string-fragmentation parameters is estimated by a nine-point variation of the Lund $a$ and $b$ parameters, combining individual variations of $\pm 0.05$ using the reweighting strategy of \cite{Bierlich:2023fmh}. 

In all cases, light-quark masses, including the one of the $\mathrm{b}$-quark, are taken to be vanishing, but quark-mass thresholds are implemented at $1.2~\mathrm{G}e\mathrm{V}$ and $4.0~\mathrm{G}e\mathrm{V}$. The showers are matched to the two-jet NLO calculation by reweighting Born-level (two-jet) events by the corresponding inclusive NLO correction and correcting the first shower branching to the full tree-level matrix element as described in \cite{Norrbin:2000uu}.
The analysis is performed with \textsc{Rivet} \cite{Bierlich:2019rhm}.

\begin{figure}[t]
    \centering
    \includegraphics[width=0.45\textwidth]{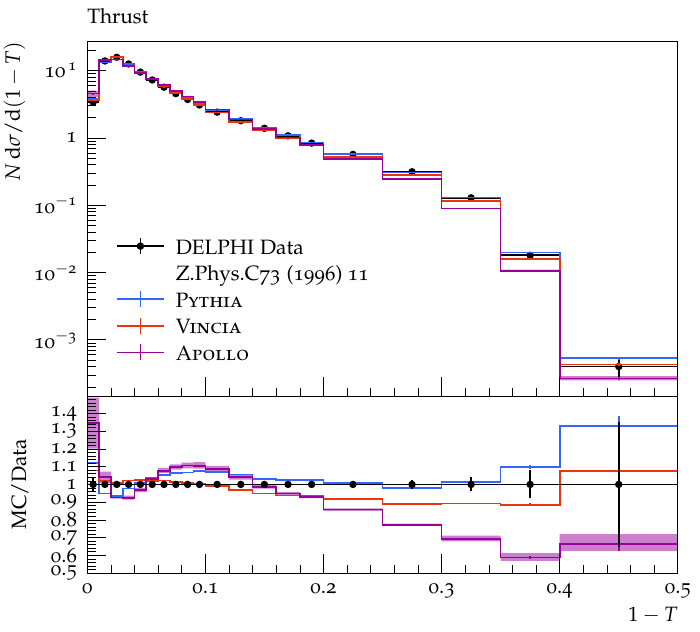}
    \includegraphics[width=0.45\textwidth]{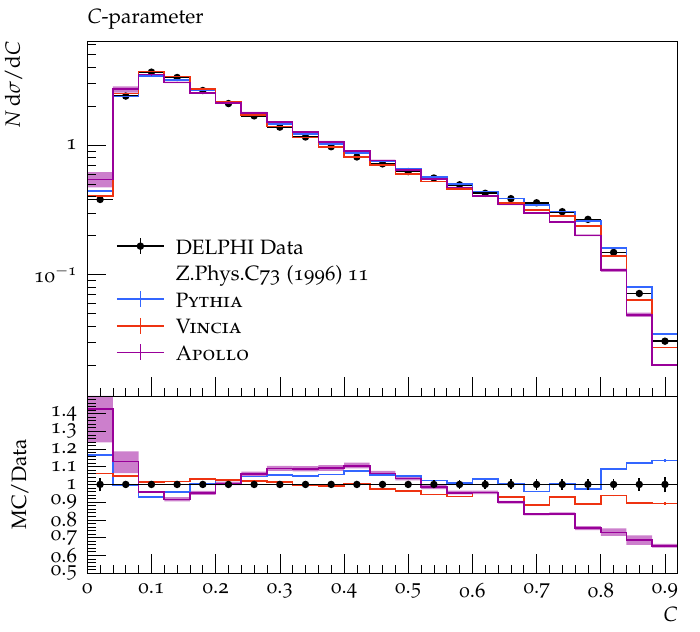}\\
    \includegraphics[width=0.45\textwidth]{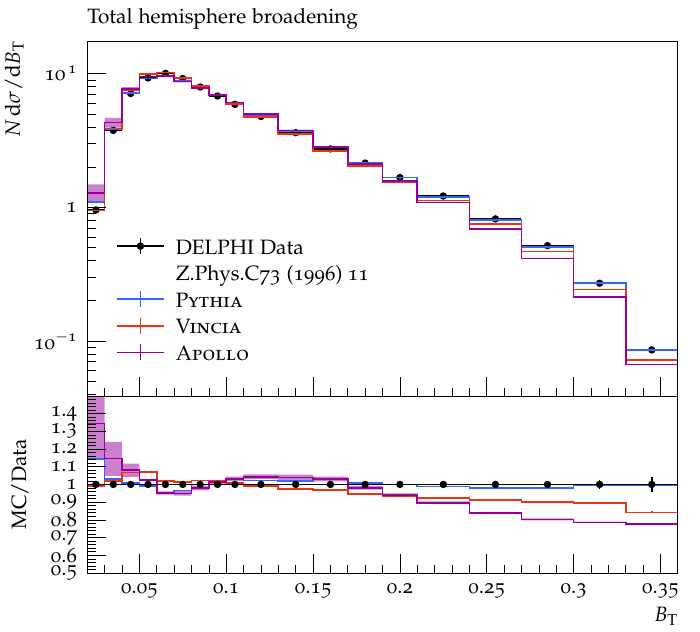}
    \includegraphics[width=0.45\textwidth]{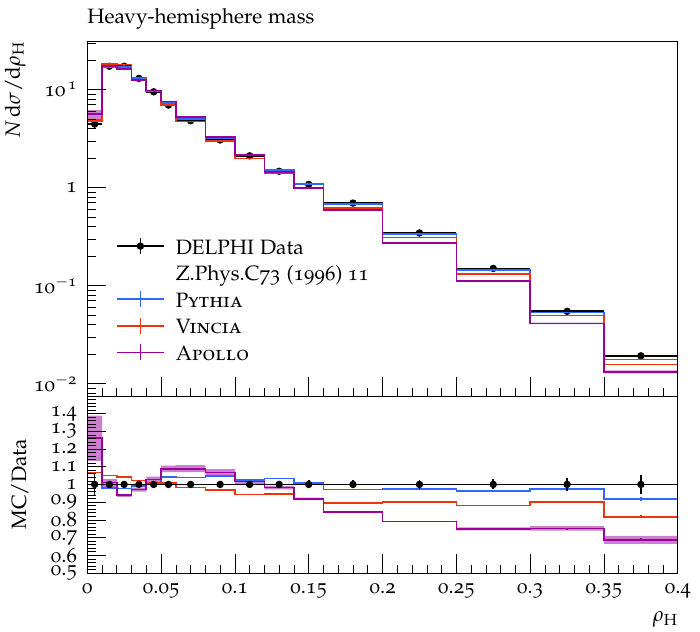}
    \caption{\apollo predictions (purple) for three-jet event-shape observables in comparison to \pythia (blue) and \vincia (red) and experimental data from the DELPHI collaboration \cite{DELPHI:1996sen}. The purple band indicates uncertainties arising from the choice of fragmentation parameters in \apollo.}
    \label{fig:compThreeJetEvtShapes}
\end{figure}

In \cref{fig:compThreeJetEvtShapes}, predictions using the \apollo shower are compared to experimental data for the three-jet event-shape observables thrust \cite{Brandt:1964sa,Farhi:1977sg}, $C$-parameter \cite{Parisi:1978eg,Donoghue:1979vi}, total hemisphere broadening \cite{Rakow:1981qn,Catani:1992jc}, and heavy-hemisphere mass \cite{Clavelli:1979md} from the DELPHI collaboration \cite{DELPHI:1996sen}. \Cref{fig:compThreeJetEvtShapes} also contains predictions obtained with \pythia and \vincia.
All parton showers acquire formal LO accuracy for these event shapes through the first-order MEC.
By the arguments of \cite{Hamilton:2023dwb}, the \apollo shower also achieves formal NNDL accuracy in these observables through the multiplicative NLO matching, whereas \pythia and \vincia achieve NDL accuracy.
In the resummation region, located centrally in the figures, good agreement between the three showers and experimental data is seen. In this region, the influence of the choice of fragmentation parameters is also minimal, indicated by the light-purple band. As expected the impact of the choice of hadronisation parameters is largest when the observable is small, i.e., towards the far left-hand side of the plots.
In the multi-particle limit, located to the right-hand side of the figures, the \apollo shower undershoots the data significantly, owing to the fact that it acquires no formal fixed-order accuracy for configurations with more than three partons, i.e., that no merging scheme is employed here. In the \pythia and \vincia showers, this is mitigated by the large value of $\alphas$ in the case of the former and the additional renormalisation-scale factors in the case of the latter.
Although not shown in the figures, it has been verified that the same applies to \apollo upon using an artificially large value of $\alphas$.

\begin{figure}[t]
    \centering
    \includegraphics[width=0.45\textwidth]{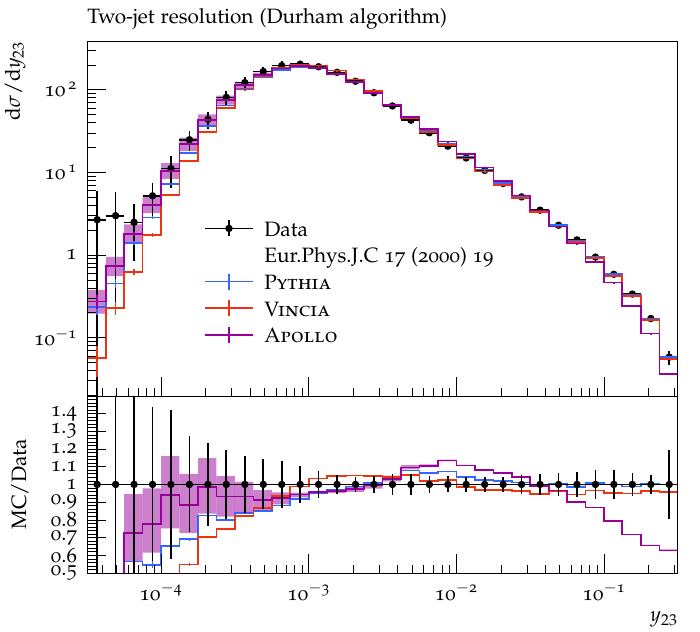}
    \includegraphics[width=0.45\textwidth]{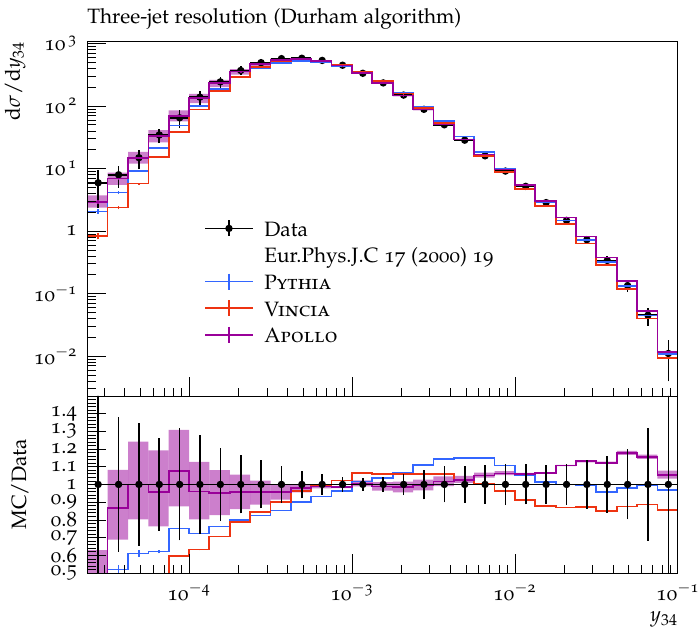}\\
    \includegraphics[width=0.45\textwidth]{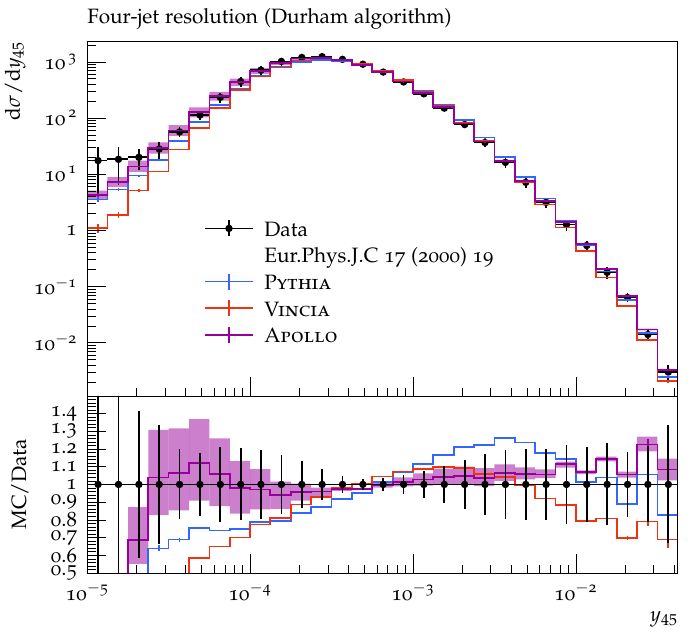}
    \includegraphics[width=0.45\textwidth]{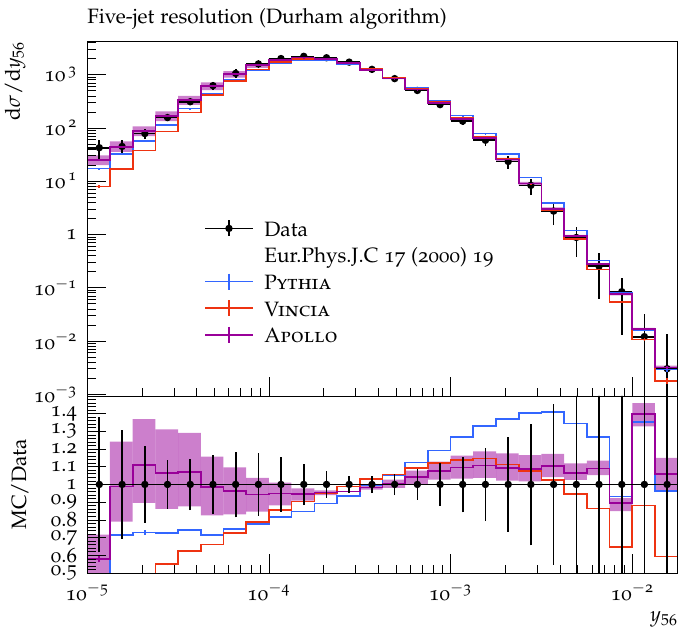}
    \caption{\apollo predictions (purple) for Durham jet-resolution scales in comparison to \pythia (blue) and \vincia (red) and experimental data from the JADE and OPAL collaborations \cite{JADE:1999zar}. The purple band indicates uncertainties arising from the choice of fragmentation parameters in \apollo.}
    \label{fig:compJetResolutions}
\end{figure}

\Cref{fig:compJetResolutions} contains \apollo predictions for jet-resolution scales for the transition to $2$, $3$, $4$, and $5$ jets in the Durham algorithm \cite{Catani:1991hj,Brown:1990nm,Brown:1991hx,Stirling:1991ds,Bethke:1991wk} in comparison to experimental data from the JADE and OPAL collaborations \cite{JADE:1999zar} and predictions from \pythia and \vincia.
By applying a first-order MEC, all showers reach LO accuracy for the two-jet transition variable $y_{23}$, while all other observables are described purely through shower branchings and therefore acquire no formal fixed-order accuracy.
It can be inferred from \cref{fig:compJetResolutions} that in the case of the two-jet resolution scale, the \apollo shower agrees fairly well with data and predictions from \pythia and \vincia for moderate values of the observable, around $y_{23} \approx 10^{-3}-5\times 10^{-2}$. Above $y_{23} = 5\times 10^{-2}$, the \apollo predictions undershoot the data quite significantly. In this region, sizeable higher-order corrections can be expected \cite{Gehrmann-DeRidder:2007vsv}.
It has been verified that the modelling of this region can significantly be improved upon choosing an artificially large value of the strong coupling, as done in \pythia and \vincia, i.e., by emulating these missing higher-order corrections.
For higher multiplicities, the agreement between the \apollo predictions and experimental data is accidentally very good, although no fixed-order corrections are included.
In all cases, the impact of hadronisation-parameter variations is the largest in the region towards the left of the figures, i.e., for small values of the observables. This aligns with expectations. 

\section{Discussion}\label{sec:Discussion}
Recent years have seen a tremendous progress on the development of new parton-shower algorithms that are consistent with resummed calculations beyond LL.
Improved parton showers with higher logarithmic accuracy have started to become available either as stand-alone codes \cite{Nagy:2020rmk,vanBeekveld:2023ivn} or as part of multi-purpose event-generation frameworks such as \sherpa \cite{Herren:2022jej,Assi:2023rbu} and \herwig \cite{Bewick:2019rbu,Bewick:2021nhc}.

In this manuscript, a new final-state shower algorithm in the \pythia event generator was presented. The new algorithm, dubbed \apollo, connects central aspects of the \vincia antenna shower with the improved transverse-recoil handling of the \alaric algorithm. Specifically, the evolution variable is defined akin to the choice in \vincia and branching kernels are constructed in analogy to NLO antenna functions. In order to employ the kinematics of \cite{Herren:2022jej,Assi:2023rbu}, antenna functions are partitioned into collinear sectors among the two antenna ends. It has been pointed out that this can be avoided in sector showers, in which the phase space is divided into non-overlapping sectors and branching kernels have to reproduce the full tree-level singularity structure of their respective sector. The resulting parton-shower algorithm differs from the \alaric shower in the evolution variable and the modelling of purely collinear parts of the branching kernels.

A key component in the design of the \apollo shower was to retain a close connection to physical matrix elements. As such, a framework for general matrix-element corrections has been outlined, as well as two independent approaches to multiplicative NLO matching, borrowing core ideas of the \powheg \cite{Nason:2004rx,Frixione:2007vw} and local analytic sector subtraction \cite{Magnea:2018hab} formalisms. The NLO matching schemes have been formulated in such a way that they retain the formal logarithmic accuracy of the \apollo shower. The validity and generality of the two NLO matching schemes has been verified in a set of numerical convergence tests.
The integration of the respective subtraction terms has been left to future work.

The consistency of the \apollo shower with NLL resummation has been demonstrated explicitly through numerical tests in the spirit of \cite{Dasgupta:2020fwr} for a wide range of global three-jet event shapes.
Preliminary predictions from \apollo have been compared to experimental measurements of event-shape observables at LEP. Good agreement between \apollo and experimental data as well as the \pythia and \vincia shower is found in the resummation region.
The agreement with data, especially in the tails of the distributions, will substantially be improved upon taking higher-order corrections to multi-jet final states into accound, e.g., via NLO merging.

A series of developments are required before the \apollo reaches a similar maturity as other shower algorithms in the \pythia framework, such as the native simple shower or \vincia. 
Most importantly, the framework has to be extended to incorporate quark-mass effects \cite{Assi:2023rbu} and initial-state radiation \cite{Bewick:2021nhc,vanBeekveld:2022zhl,vanBeekveld:2022ukn,vanBeekveld:2023chs}. 
While the former is needed for an accurate description of heavy-quark fragmentation, the latter is essential for the application to hadron-hadron and lepton-hadron collisions. Subsequently, the \apollo shower can be fully embedded into \pythia's internal merging machinery, allowing the use of the CKKW-L, UMEPS, and UNLOPS schemes \cite{Lonnblad:2011xx,Lonnblad:2012ix,Lonnblad:2012ng}.
Especially the inclusion of the NLO three-jet correction in lepton collisions is necessary to derive a consistent tune. 
Concerning the logarithmic accuracy of the \apollo shower, both the inclusion of spin correlations \cite{Collins:1987cp,Knowles:1987cu,Knowles:1988hu} and subleading-colour corrections \cite{Hamilton:2020rcu,Gustafson:1992uh} are necessary to claim NLL accuracy across the board. 
Finally, the multiplicative matching methods outlined here are formulated in such a way that they facilitate extensions to next-to-next-to-leading order \cite{Campbell:2021svd}.

\acknowledgments
The author would like to thank Stefan Höche and Peter Skands for many helpful discussions and comments on the manuscript.
Computations were carried out on the PLEIADES cluster at the University of Wuppertal, supported by the Deutsche Forschungsgemeinschaft (DFG, grant No. INST 218/78-1 FUGG) and the Bundesministerium für Bildung und Forschung (BMBF).

\appendix
\section{Explicit expressions for matrix-element-corrected branching kernels}\label{sec:MECBranchingKernels}
The tree-level matrix element for the off-shell photon decay $\upgamma^*\to \q\qbar\g$ is given by
\begin{equation}
    \vert \mc{M}_{\gamma\to\q\qbar\g}^0\vert^2 = 2\CF\gs^2\, \left[\frac{2s_{ik}}{s_{ij}s_{jk}} + \frac{s_{jk}}{s_{ij}s_{ijk}} + \frac{s_{ij}}{s_{jk}s_{ijk}}\right]\, \vert\mc{M}_{\gamma^*\to\q\qbar}^0\vert^2 \,.
\end{equation}
The matrix-element-corrected branching kernel \cref{eq:Pqg} then reads
\begin{equation}
    P_{\upgamma^*\to\q\g,\qbar}(p_i,p_j,p_k;n_j) = \frac{2s_{ik}(p_in_j)}{s_{ij}(s_{jk}(p_in_j)+s_{ij}(p_kn_j))} + \frac{s_{jk}}{s_{ij}s_{ijk}} \, .
\end{equation}

The tree-level matrix element for Higgs decays into a massless quark pair $\H\to\q\qbar\g$ is given by
\begin{equation}
    \vert \mc{M}_{\H\to\q\qbar\g}^0\vert^2 = 2\CF\gs^2\, \left[\frac{2s_{ik}}{s_{ij}s_{jk}} + \frac{s_{jk}}{s_{ij}s_{ijk}} + \frac{s_{ij}}{s_{jk}s_{ijk}} + \frac{2}{s_{ijk}}\right]\, \vert\mc{M}_{\H\to\q\qbar}\vert^2 \,.
\end{equation}
The matrix-element-corrected branching kernel \cref{eq:Pqg} can then be taken as
\begin{equation}
    P_{\H\to\q\g,\qbar}(p_i,p_j,p_k;n_j) = \frac{2s_{ik}(p_in_j)}{s_{ij}(s_{jk}(p_in_j)+s_{ij}(p_kn_j))} + \frac{s_{jk}}{s_{ij}s_{ijk}} + \frac{1}{s_{ijk}}\, .
\end{equation}

The tree-level matrix element for Higgs decays into three gluons $\gamma^*\to \g\g\g$ in the HEFT is given by
\begin{equation}
    \vert \mc{M}_{\H\to\g\g\g}^0\vert^2 = 2\CF\gs^2\, \left[\frac{2s_{ik}}{s_{ij}s_{jk}} + \frac{s_{jk}s_{ik}}{s_{ij}s_{ijk}^2} + \frac{s_{ij}s_{ik}}{s_{jk}s_{ijk}^2} + \frac{8}{3s_{ijk}} + (i \leftrightarrow j) + (j\leftrightarrow k) \right]\, \vert\mc{M}_{\H\to\g\g}\vert^2 \,.
\end{equation}
The tree-level matrix element for Higgs decays into three gluons $\gamma^*\to \q\qbar\g$ in the HEFT is given by
\begin{equation}
    \vert \mc{M}_{\H\to\q\qbar\g}^0\vert^2 = 2\CF\gs^2\, \left[\frac{1}{s_{ij}} - \frac{2s_{jk}s_{ik}}{s_{ij}s_{ijk}^2} - \frac{2s_{ijk} - s_{ij}}{s_{ijk}^2}\right]\, \vert\mc{M}_{\H\to\g\g}\vert^2 \,.
\end{equation}
The respective matrix-element-corrected branching kernel \cref{eq:Pgg} and \cref{eq:Pqq} can then be constructed to be
\begin{align}
    P_{\H\to\g\g,\g}(p_i,p_j,p_k;n_j) &= \frac{2s_{ik}(p_in_j)}{s_{ij}(s_{jk}(p_in_j)+s_{ij}(p_kn_j))} + \frac{s_{jk}s_{ik}}{s_{ij}s_{ijk}^2} + \frac{s_{ij}s_{ik}}{s_{jk}s_{ijk}^2} + \frac{8}{6s_{ijk}} \, , \\
    P_{\H\to\q\qbar,\g}(p_i,p_j,p_k;n_j) &= \frac{1}{s_{ij}} - \frac{2s_{jk}s_{ik}}{s_{ij}s_{ijk}^2} - \frac{2s_{ijk} - s_{ij}}{s_{ijk}^2} \, .
\end{align}

\section{Branching kinematics and inverse construction}\label{app:kinematics}
For gluon emissions, the kinematics are constructed in the event frame as follows,
\begin{equation}
\begin{split}
    p_i^\mu &= z \tilde{p}_{ij}^\mu \, , \\
    p_j^\mu &= a\tilde{p}_{ij}^\mu + b\tilde{K}^\mu + p_\perp^\mu\, , \\
    K^\mu &= (1-z-a)\tilde{p}_{ij}^\mu + (1-b)\tilde{K}^\mu - p_\perp^\mu \, ,
\end{split}
\end{equation}
with a subsequent boost of all momenta in the event into the rest frame of $K^\mu$.
The parameters of the map are given in terms of
\begin{equation}
    z = \frac{p_in_j}{(p_i+p_j)n_j} \, , \quad \text{and}\quad  \kappa = \frac{\tilde{K}^2}{2\tilde{p}_{ij}\tilde{K}} 
\end{equation}
as follows
\begin{equation}
    \quad a = (1-b)(1-z)-2b\kappa \, , \quad b = \frac{1}{z}\frac{s_{ij}}{2\tilde{p}_{ij}\tilde{K}} \, .
\end{equation}
The inverse of the kinematics are constructed in terms of $z$ as
\begin{equation}
\begin{split}
    \tilde{p}_{ij}^\mu &= \frac{1}{z}p_i^\mu \, , \\
    \tilde{K}^\mu &= K^\mu + p_j^\mu - \frac{1-z}{z}p_i^\mu \, ,
\end{split}
\end{equation}
with a subsequent boost into the rest frame of $\tilde{K}$.

For gluon splittings, the kinematics are constructed in the event frame as
\begin{equation}
\begin{split}
    p_i^\mu &= a_i\tilde{p}_i^\mu + b_i\tilde{K}^\mu + p_\perp^\mu \, , \\
    p_j^\mu &= a_j\tilde{p}_i^\mu + b_j\tilde{K}^\mu - p_\perp^\mu \, ,\\
    K^\mu &= (1-a_i-a_j)\tilde{p}_{ij} + (1-b_i-b_j)\tilde{K}^\mu \, ,
\end{split}
\end{equation}
with a subsequent boost into the rest frame of $K^\mu$. 
The parameters of the map are given in terms of
\begin{equation}
  \bar{z} = \frac{p_iK}{(p_i+p_j)K} \, ,\quad y = \frac{p_ip_j}{p_ip_j+(p_i+p_j)K} \, ,
\end{equation}
and $\kappa$ by
\begin{equation}
\begin{split}
    a_i &= \frac{1}{2(1+\kappa)}\left(\bar{z}+(1-\bar{z})y+(1+2\kappa)\frac{(1-y)(y-\bar{z}(1+y))+2y\kappa}{\sqrt{(1-y)^2-4y\kappa}} \right) \, ,\\
    a_j &= \frac{1}{2(1+\kappa)}\left((1-\bar{z})+\bar{z}y-(1+2\kappa)\frac{(1-y)(1-\bar{z}(1+y))-2y\kappa}{\sqrt{(1-y)^2-4y\kappa}}\right) \, ,\\
    b_i &= \frac{1}{2(1+\kappa)}\left(\bar{z}+(1-\bar{z})y-\frac{(1-y)(y-\bar{z}(1+y))+2y\kappa}{\sqrt{(1-y)^2-4y\kappa}} \right) \, ,\\
    b_j &= \frac{1}{2(1+\kappa)}\left((1-\bar{z})+\bar{z}y+\frac{(1-y)(1-\bar{z}(1+y))-2y\kappa}{\sqrt{(1-y)^2-4y\kappa}}\right) \, ,
\end{split}
\end{equation}
The inverse of the gluon-splitting kinematics is given by
\begin{equation}
\begin{split}
    \tilde{p}_{ij}^\mu &= \tilde{a}_i (p_i^\mu+p_j^\mu) + \tilde{b}_i K^\mu \, ,\\
    \tilde{K}^\mu &= (1-\tilde{a}_i)(p_i^\mu+p_j^\mu) + (1-\tilde{b}_i)K^\mu \, ,
\label{eq:splittingKinematicsInverse}
\end{split}
\end{equation}
with a subsequent boost into the rest frame of $\tilde{K}$. The parameters of the inverse map are determined by $s_{ij}$ and $y$ as
\begin{equation}
\begin{split}
    \tilde{a}_i &= \frac{s_{ij}\tilde{\rho}-s_{ij}(1-y)-2yK^2}{2(s_{ij}+y K^2)\tilde{\rho}} \, ,\\
    \tilde{b}_i &= \frac{s_{ij}\tilde{\rho}+s_{ij}(1+y)}{2(s_{ij}+y K^2)\tilde{\rho}} \, ,\\
    \tilde{\rho} &= \sqrt{(1-y)^2-\frac{4yK^2}{s_{ij}}} \, .
\end{split}    
\end{equation}

\bibliography{main.bib}

\end{document}